\documentclass[prl, twocolumn,superscriptaddress,preprintnumbers,amsmath,amssymb]{revtex4}
\usepackage{dcolumn}
\usepackage{bm}
\usepackage{amsmath,amsfonts,amssymb,ulem}
\usepackage{epstopdf}
\usepackage{xcolor}
\usepackage{color}
\usepackage{wasysym}

\usepackage{graphicx}
\usepackage[colorlinks = true,
            linkcolor = blue,
            urlcolor  = blue,
            citecolor = blue,
            anchorcolor = blue]{hyperref}

\begin{document}

\title{Energetic spin-polarized proton beams from two-stage coherent acceleration in laser-driven plasma}

\author{Zheng Gong}
\altaffiliation{Present address: Max Planck Institute for Nuclear Physics, Saupfercheckweg 1, D-69117 Heidelberg, Germany}
\affiliation{SKLNPT, KLHEDP, CAPT and School of Physics, Peking University, Beijing 100871, China}

\author{Yinren Shou}
\affiliation{SKLNPT, KLHEDP, CAPT and School of Physics, Peking University, Beijing 100871, China}
\author{Yuhui Tang}
\affiliation{SKLNPT, KLHEDP, CAPT and School of Physics, Peking University, Beijing 100871, China}

\author{Xueqing Yan}
\email{x.yan@pku.edu.cn}
\affiliation{SKLNPT, KLHEDP, CAPT and School of Physics, Peking University, Beijing 100871, China}
\affiliation{CICEO, Shanxi University, Taiyuan, Shanxi 030006, China}


\date{\today}
\begin{abstract}

We propose a scheme to overcome the great challenge of polarization loss in spin-polarized ion acceleration. When a petawatt laser pulse penetrates through a compound plasma target consisting of a double layer slab and prepolarized hydrogen halide gas, a strong forward moving quasistatic longitudinal electric field is constructed by the self-generated laser-driven plasma. This field with a varying drift velocity efficiently boosts the prepolarized protons via a two-stage coherent acceleration process. Its merit is not only achieving a highly energetic beam but also eliminating the undesired polarization loss of the accelerated protons. We study the proton dynamics via Hamiltonian analyses, specifically deriving the threshold of triggering the two-stage coherent acceleration. To confirm the theoretical predictions, we perform three-dimensional PIC simulations, where unprecedented proton beams with energy approximating half GeV and polarization ratio $\sim$ 94\% are obtained. 

\end{abstract}


\maketitle

Spin is an essential intrinsic property of ions~\cite{griffiths2018introduction,mane2005_review}. Energetic spin-polarized proton (SPP) beams are extensively used in fundamental physics~\cite{new_physics_from_spin,Rep_Pro_Phy_2019} in exploring internal structures of nucleon~\cite{Ji_2017_PRL_nucleon_spin,gluon_spin_contribution_to_the_proton_spin}, nonperturbative quantum chromodynamics~\cite{investigate_QCD,Yang_2017_QCDsim,Alexandrou2017_QCDsim}, parity violating spin asymmetry in polarized proton colliders~\cite{parity_violating_spin_asymmetry_PRL}, and exotic phenomena within or beyond the standard model~\cite{np_2020_beyond_standard}. In nuclear physics, the SPP acts as a probe to measure the cross section of nucleus interaction~\cite{rosen1967polarized_Science,Tojo_PRL_2002,Allgower_PRD_2002}, such as electron capture involved with the giant dipole resonance~\cite{glavish1972giant_GDR} and photon emission from nucleon bremsstrahlung~\cite{kitching1986polarized_pp_bremsstrahlung}. Additionally, SPPs have also been pursued in industrial applications, e.g., electrochemical membrane optimization~\cite{kee2013modeling} and highly sensitive biomedical imaging~\cite{zimmer2016polarized}. Previously, energetic SPP beams were provided by traditional accelerators~\cite{bai2006polarized} equipped with the corkscrew-like magnets, i.e., \textit{Siberian snakes}~\cite{derbenev1975acceleration_sebrian_snake}, to minimize proton polarization loss. The defect of such accelerators is too large in scale and budget. Therefore, an alternative compact and economical design for producing energetic SPP beams is highly desired.

Cutting-edge facilities based on the frontier optical technology~\cite{CPA_1985compression,Mourou_etal_2006} can realize pulse intensity far beyond $10^{20}$W/cm$^2$~\cite{danson2019petawatt_laser}, which enables low-cost and efficient laser plasma accelerators with the gradient over 100GeV/m~\cite{tajima1979,Esarey_2009_RMP,macchi_2013_RMP}. On the other side, owing to the ultraviolet photodissociation method~\cite{rakitzis2003spin_science,Sofikitis2008Laser_JCP,sofikitis2017spin_PRL,boulogiannis2019spin}, nuclear spin polarized hydrogen densities extended to $10^{19}\mathrm{cm}^{-3}$ with lifetimes near 10 ns have been achieved experimentally~\cite{sofikitis2018spin_PRL}. Encouraged by the above two breakthroughs, there is increasing interest in high efficiency laser-driven particle acceleration~~\cite{wen2019polarized,wu2019polarized_NJP,wu2019polarized_PRE} and spin-polarized inertial confinement fusion~\cite{Hu_2020_PRE} in prepolarized plasma. However, up to now, only a few schemes have been proposed to accelerate polarized protons~\cite{H2019Polarized,Buscher_2019,jin2020spin} and there is still a lack of an insightful understanding of the coulping effect between proton dynamics and collective plasma phenomena. As a result, the generated SPP beam is in low quality, with energy $\sim$ 50 MeV and a polarization ratio $<$ 80\%~\cite{jin2020spin}, which inevitably limits the relevant applications requiring an energetic and highly polarized proton beam~\cite{Rep_Pro_Phy_2019,Tojo_PRL_2002,kitching1986polarized_pp_bremsstrahlung}. 

In this paper, we report an approach to generate highly energetic SPP beams whose maximum energy is close to half GeV and polarization ratio is as high as 94\% by utilizing a petawatt laser. When the intense pulse propagates through a compound plasma target, a strong quasi static longitudinal electric field (QSLEF) with a varying drift velocity repeatedly accelerates the polarized protons through a two-stage coherent process. Earlier, the protons are swiftly reflected by the forward moving QSLEF to arrive at a moderate velocity, while their spin polarization is largely preserved on account of the negligible net accumulation of spin modulation induced by the oscillating laser magnetic field. Later, as the drift QSLEF moves faster than these pre-accelerated protons, the protons will be caught, trapped, and reflected again by the drift QSLEF to reach higher energy.  Meanwhile, a vortex plasma magnetic field, issuing in the uncompensated transverse spin precession, merely leads to a minor polarization decrease $\sim$ 6\% for the generated SPP beam. Because net spin precession occurs within a short duration in the second stage, the realized energetic SPP beam still maintains a high polarization ratio $\sim$ 94\%, which would significantly facilitate the development of multiple branches of physics.

\begin{figure}[t]
\includegraphics[width=0.99\columnwidth]{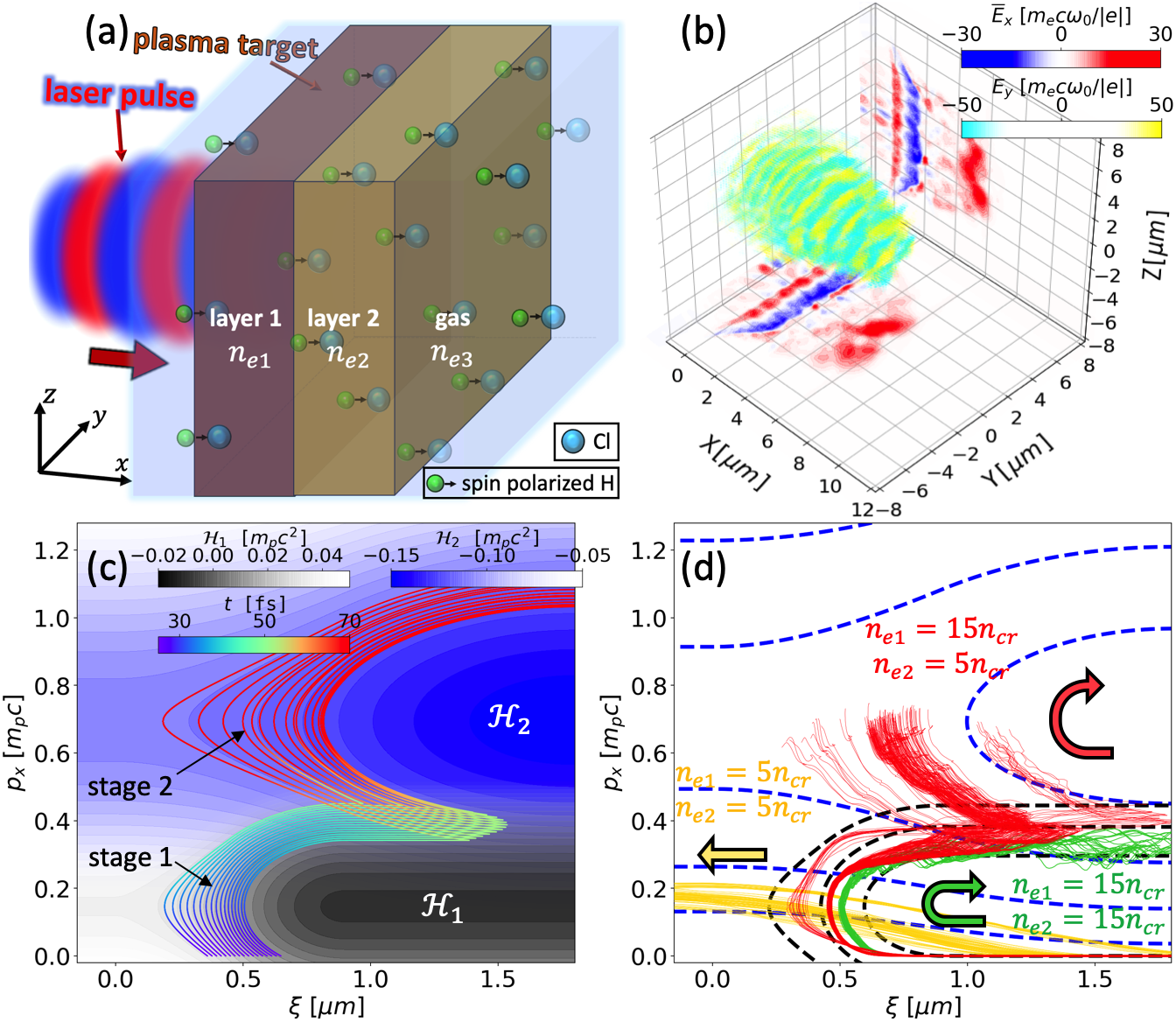}
\caption{(a) Schematic diagram of the compound plasma target irradiated by a laser pulse. (b) the volume rendering refers to spatial distribution of field $E_y$ while the projection exhibits $\overline{E}_x$. (c) theoretically predicted proton trajectories in $(\xi,p_x)$ space, where the rainbow color denotes the time and the black (blue) colormap displays the distribution of $\mathcal{H}_1$ ($\mathcal{H}_2$). (d) typical proton trajectories extracted from 3D PIC simulations.}
\label{fig:schematic}
\end{figure}

The SPP dynamics is studied with the three-dimensional (3D) particle-in-cell (PIC) code Epoch~\cite{arber2015contemporary}. The simulation domain 15$\mu m$$\times$20$\mu m$$\times$20$\mu m$ is discretized into 600$\times$400$\times$400 grid cells. The compound target (s. Fig.~\ref{fig:schematic}a) is a double layer slab of near-critical density carbon nanotube foams~\cite{ma2007directly} mixed with the prepolarized HCl gas~\cite{sofikitis2018spin_PRL}. The electron densities $n_e$ of the first and second layer are $n_{e1}=15n_{cr}$ and $n_{e2}=5n_{cr}$, respectively. Here $n_{cr} \equiv m_e \omega_l^2 / 4 \pi e^2$ is the classical critical density that determines plasma opacity in nonrelativistic regime for a laser with frequency $\omega_l$. $m_e$ and $e$ are the electron mass and charge. The first (second) layer target is placed at $0\leqslant x[\mu m]<3$ ($3\leqslant x[\mu m]\leqslant 6$) where 40 particles per cell are chosen. The density profile of the hydrogen halides gas is a trapezoid with a flat top at $0\leqslant x[\mu m]\leqslant 9$ and $1 \mu m$ ramp on both sides, where each cell is filled with 20 particles. The electron density of the prepolarized HCl is $n_{e3}=0.8 n_{cr}$, which is equivalent to experimentally accessible hydrogen density $4.9\times 10^{19} \mathrm{cm}^{-3}$~\cite{sofikitis2018spin_PRL}. The circularly polarized laser pulse of Gaussian longitudinal envelope with intensity $a_0 \approx 70 $, wavelength $\lambda = 1$ $\mu$m, spot size $5.8 \mu m$ and duration $30$ fs in FWHM, is focused at the plane $x=0\mu m$. Considering that such an ultraintense laser usually have a few hundred femtoseconds pre-pulse with relativistic intensities before the main pulse, we perform additional 1D PIC simulations to ensure that a pre-pulse with a duration of 400 fs and intensities up to $\sim 10^{18}\mathrm{W/cm}^2$ has insignificant impact on the initial condition of plasma target. Therefore, the scheme presented here for producing energetic SPP beams is valid and robust for the ultraintense pulse with a relatively high contrast ratio.


When the ultraintense pulse stably propagates inside the plasma, the drift QSLEF is sustained by the charge separation accumulated at the front edge of laser pulse as shown in Fig.~\ref{fig:schematic}b, where the projection displays the distribution of longitudinal electric field $\overline{E}_x$ averaged over two laser periods at section $z=0$ and $y=0$. The drift QSLEF $\overline{E}_x$, the source of accelerated proton energies, moves along the laser propagation, as in previous results~\cite{zhang2007multistaged,robinson2008radiation,robinson2009relativistically,naumova2009hole,weng2012ultra}. It is convenient to characterize the proton motion in the moving frame of the drift QSLEF in one dimensional circumstance. Here, we adopt a local constant approximation, where the whole acceleration process is divided into several separate stages and in each of them the drift QSLEF is assumed to be independent on time. Thus, the properties of QSLEF are described by drift velocity $v_i$, relative coordinate $\xi=x-v_it$, field strength $E_i(\xi)$ ($\partial E_i/\partial t=0$), and electric potential $\varphi_i(\xi)$ ($\partial \varphi_i/\partial t=0$), where the subscript \textit{i} denotes the ordinal number of stages. At each stage, after reformulating the proton dynamic equation in $(\xi,p)$ space, a Hamiltonian is presented as $\mathcal{H}_i(\xi,p)=c\sqrt{m_p^2c^2+p^2}-v_ip+|e|\varphi_i(\xi)$~\cite{shen2007bubble,gong2020proton}, where $c$ the speed of light, $m_p$ the proton mass and $\xi_d$ ($\xi_u$) the downstream (upstream) boundary of electric potential. Given the conservation of Hamiltonian $\mathcal{H}_i$ between points $(\xi_d,p^{r\pm}_i)$ and $(\xi_u,p_i)$ along the separatrix, where $p_i=m_pc\beta_i/\sqrt{1-\beta_i^2}$ and $\beta_i=v_i/c$, the upper and lower limit momenta $p^{r\pm}_i$ of protons along the contour of $\mathcal{H}_i(\xi,p)=\mathcal{H}_i(\xi_u,p_i)$ at the $i$-th stage can be derived as $p^{r\pm}_i/(m_pc)=(\beta_i\mathcal{A}_i\pm\sqrt{\mathcal{A}_i^2+\beta_i^2-1})/(1-\beta_i^2)$, where $\mathcal{A}_i=|e|\varphi_i(\xi_u)/(m_pc^2)+1/\gamma_i$ and $\gamma_i=1/\sqrt{1-\beta_i^2}$.

\begin{figure}[t]
\includegraphics[width=0.99\columnwidth]{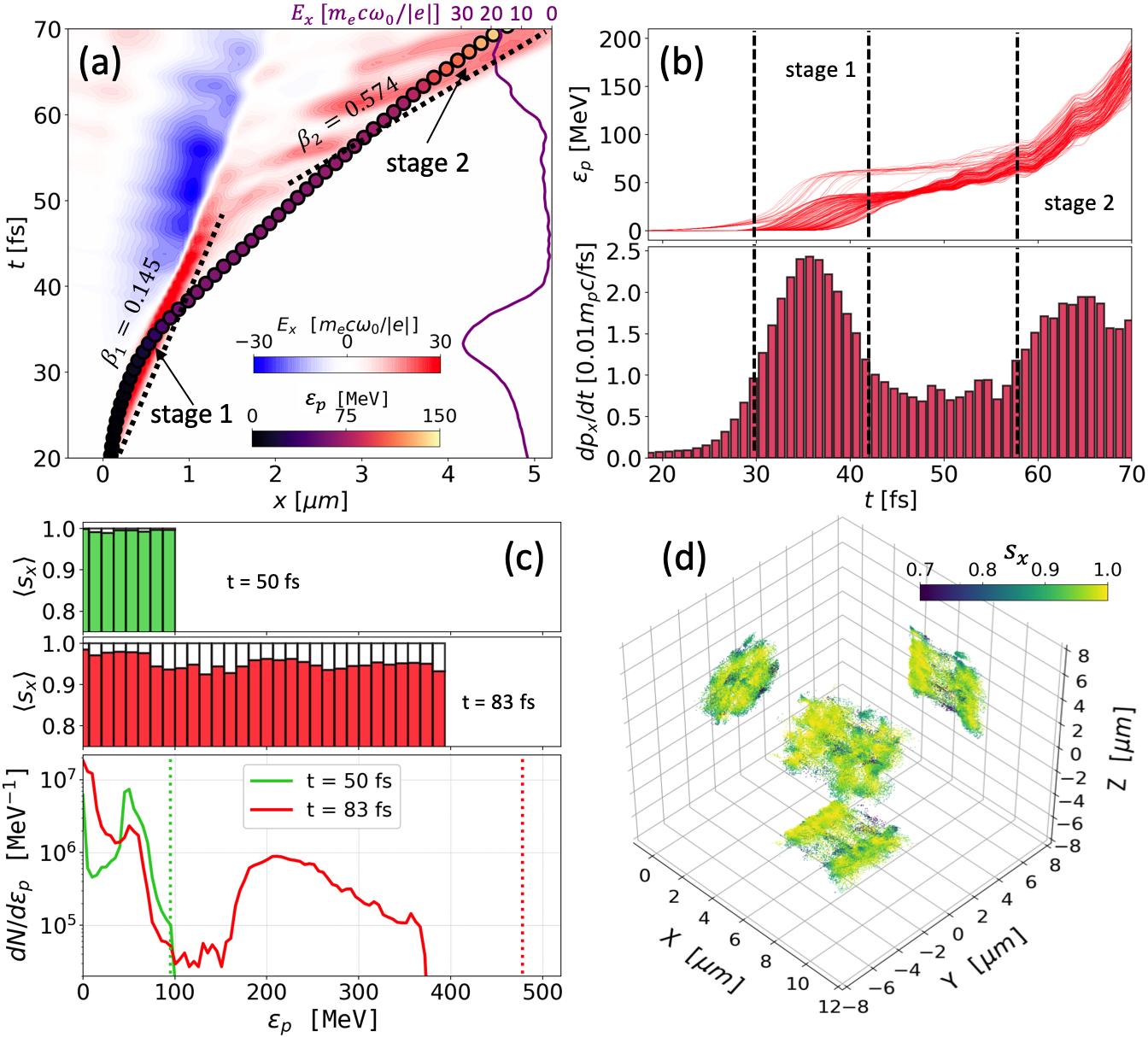}
\caption{(a) a proton trajectory (rendered with magma color for its energy $\varepsilon_p$) inside the evolution of the on-axis accelerating field $E_x(y=0,z=0)$. (b) the time evolution of proton energy $\varepsilon_p$ and momentum differential $dp_x/dt$. (c) the proton energy spectra (in the lower panel) and the spin polarization ratio for different energy 
range (in the upper panel), where the panels share the same horizontal axis. (d) the spatial distribution of proton spin $s_x$ at $t=83$ fs.}
\label{fig2:detail_acc}
\end{figure}

In order to realize the separate multiple stage proton acceleration, the connection between stages $i$ and $i+1$ requires that the pre-accelerated protons by the stage-$i$ can be caught up, trapped, and reaccelerated by the later stage-$i+1$. From mathematical aspect, it is equivalent to the contours of $\mathcal{H}_{i+1}(\xi,p)=\mathcal{H}_{i+1}(\xi_u,p_{i+1})$ and $\mathcal{H}_{i}(\xi,p)=\mathcal{H}_{i}(\xi_u,p_i)$ intersecting with each other. Consequently, the condition for successfully coupling these two stages are expressed as
\begin{eqnarray}\label{eq:intersect}
p^{r+}_{i}-p^{r-}_{i+1}>0, 
\end{eqnarray}
where the value of $p^{r\pm}_i$ can be determined once $\beta_{i}$ and $\varphi_i(\xi_u)$ are given. The solution of Eq.\eqref{eq:intersect} is analytically derived as $\beta_{i+1}<\beta^*$ where

\begin{eqnarray}\label{eq:threshold_beta}
\beta^*\equiv\frac{-\Tilde{\varphi}_{i+1}\Tilde{p}+\Tilde{p}\Tilde{\gamma}+\sqrt{\Tilde{\varphi}_{i+1}[2\Tilde{\gamma}-\Tilde{\varphi}_{i+1}]}}{\Tilde{\gamma}^2}.
\end{eqnarray}
Here $\Tilde{\varphi}_{i+1}=|e|\varphi_{i+1}(\xi_u)/(m_pc^2)$, $\Tilde{p}= p^{r+}_i/(m_pc)$, and $\Tilde{\gamma}=\sqrt{1+\Tilde{p}^2}$ are utilized. The detailed illustration of Eq.\eqref{eq:intersect} and the derivation for obtaining Eq.\eqref{eq:threshold_beta} are given in the Appendix. The rainbow lines in Fig.~\ref{fig:schematic}c exhibit the theoretical proton trajectories in $(\xi,p)$ space under a two-stage coherent acceleration, where the black (blue) contour represents the distribution of Hamiltonian $\mathcal{H}_1$ ($\mathcal{H}_2$) at the first (second) stage. For $\mathcal{H}_1(\xi,p)$, given that the parameters $\beta_1=0.145$ and $\Tilde{\varphi}_1\approx\Tilde{\varphi}_2\approx 0.0462$ are calculated from the moving longitudinal electric field $E_x$ based on the 3D PIC simulation as shown in Fig.~\ref{fig2:detail_acc}(a), one can find $\Tilde{p}=0.466$ and $\Tilde{\gamma}=1.103$. Substituting these values into Eq.~\eqref{eq:threshold_beta}, we arrive at $\beta^*\approx0.664>\beta_2=0.574$, which indicates the success of connecting these two separate acceleration stages. 

The representative proton trajectories within time $20<t [\mathrm{fs}]<70$ extracted from 3D PIC simulations are shown as red lines in Fig.~\ref{fig:schematic}d, where the background black (blue) dashed lines denote the contour of $\mathcal{H}_1$ ($\mathcal{H}_2$), the same as that in Fig.~\ref{fig:schematic}c. The protons (in red) are reaccelereted to a momentum $p_x\approx 0.7 m_pc$ at $t= 70$ fs in the second Hamiltonian $\mathcal{H}_2(\xi,p)$ after being preaccelerated to $p_x\approx 0.3 m_pc$ in $\mathcal{H}_1(\xi,p)$. For comparison, green trajectories standing for the case of a uniform target with $n_{e1}=n_{e2}=15n_{cr}$ demonstrate that the protons merely experience the first stage of reflection in $\mathcal{H}_1(\xi,p)$. Additionally, the criterion of an initially resting proton being trapped in the Hamiltonian $\mathcal{H}_2(\xi,p)$ is calculated as $\beta^*\approx0.300<\beta_2=0.574$, and thus the yellow trajectories representing the case of $n_{e1}=n_{e2}=5n_{cr}$ exhibit that the protons quickly sliding away in $\mathcal{H}_2(\xi,p)$ are not trapped by the faster drift QSLEF.

To further understand the two-stage process, we visualize a typical proton trajectory in time resolved $(x,t)$ coordinate (s. Fig.~\ref{fig2:detail_acc}a), where the background color denotes the strength of electric field $E_x(y=0,z=0)$ at the central axis. When the laser bores a hole on the first layer slab, the proton is reflected by the laser pulse in the first stage to have a longitudinal velocity $\beta_x\approx 0.330>\beta_1$. After the laser pulse penetrates through the first layer and irradiates on the second layer slab, a faster drift QSLEF $E_x$ with velocity $\beta_2=0.574>\beta_x$ is generated to catch up with the protons and further accelerate them. The second stage of acceleration mainly occurs at $3<x[\mu m]\lesssim 5$ which is in accordance with the location of the second layer slab. Another pronounced signal of the discontinuous two-stage acceleration is the purple line which profiles the strength of accelerating field $E_x$ imposed on the proton during $20<t[\mathrm{fs}]<70$. The upper panel of Fig.~\ref{fig2:detail_acc}b shows that the proton energies $\varepsilon_p$ increase predominantly during $30\lesssim t[\mathrm{fs}]\lesssim 40$ and $60\lesssim t[\mathrm{fs}]\lesssim 70$, and the maximum acceleration ratio $d\varepsilon_p/dt$ is up to 10 MeV/fs. The momentum differential $dp_x/dt$ averaged over the whole typical protons is illustrated in the lower panel of Fig.~\ref{fig2:detail_acc}b and the maximum accelerating gradient is over 60 TeV/m. One advantage of this mechanism is the energy enhancement induced by the second stage acceleration, which is identified by the proton energy spectra at the end of two stage ($t=50$ and 83 fs) in the lower panel of Fig.~\ref{fig2:detail_acc}c, where the vertical dotted lines refer to the theoretically predicted maximum energy $\varepsilon_p^\mathrm{max}=[m_p^2c^2+(p^{r+})^2]^{1/2}=94.7$ and 478 MeV in each stage. It is worth emphasizing that the key point in this mechanism is the twice occurrence of spatial coherence between the SPPs and the drift QSLEF inside these two plasma slabs. This is far different from energetic SPP bunches driven by magnetic vortex acceleration~\cite{jin2020spin}, where the proton energy is predominantly obtained when the laser pulse exits the rear surface of gas targets.

For the purpose of unveiling the coupling between the spin and laser-plasma effect, by utilizing Thomas-BMT equation~\cite{thomas1927kinematics,bargmann1959precession}, we can characterize the proton spin dynamics as $d\mathbf{s}/dt=\mathbf{\Omega}\times\mathbf{s}$ where
\begin{small}\begin{eqnarray}
\mathbf{\Omega}=\frac{e}{m_pc}[\frac{a\gamma+1}{\gamma}\mathbf{B}-\frac{a\gamma}{\gamma+1}(\bm{\beta}\cdot\mathbf{B})\bm{\beta}-\frac{a\gamma+a+1}{\gamma+1}\bm{\beta}\times\mathbf{E}].
\label{T-BMT}
\end{eqnarray}
\end{small}Here, $a\approx1.7928$ is the anomalous magnetic moment for proton. The spatial distribution of spin $s_x$ of protons with energy $\varepsilon_p>50$ MeV manifests the highly polarization of the generated energetic SPP beam (s. Fig.~\ref{fig2:detail_acc}d). The upper panel of Fig.~\ref{fig2:detail_acc}c presents the polarization ratio $\left<S_x\right>$ averaged over the whole protons within each energy bin. At the end of the first stage $t=50$ fs, the averaged spin polarization ratio is $\left<S_x\right> \approx 0.994$, whilst at $t=83$ fs the ratio $\left<S_x\right> \approx 0.946$. The polarization loss $1-\left<S_x\right>\approx 0.054$ at $t=83$ fs is higher than that at $t=50$ fs. Detailed tracking of proton spin is illustrated below to explain the reason. 

\begin{figure}[t]
\includegraphics[width=0.99\columnwidth]{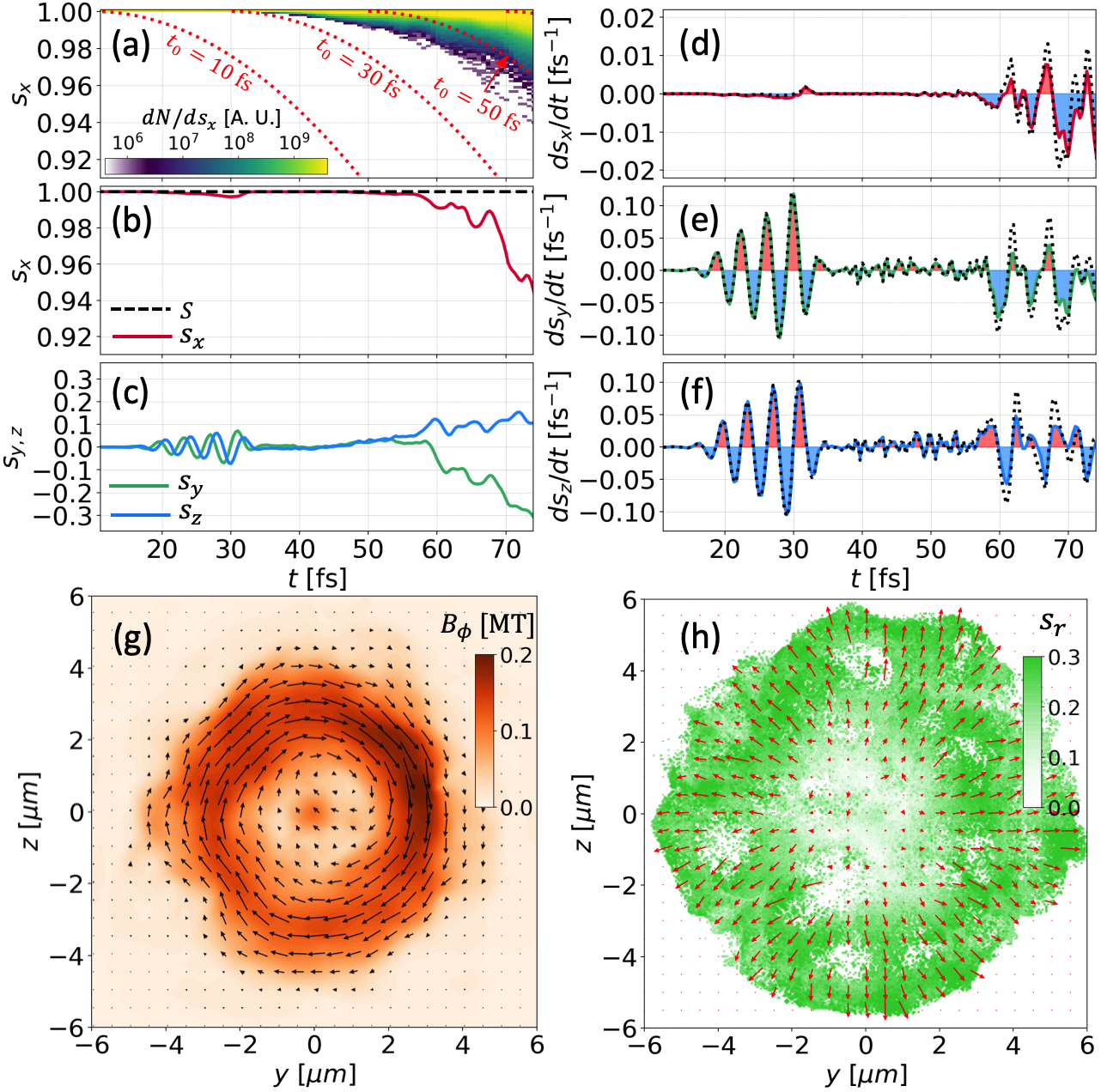}
\caption{(a) Time-resolve proton spin distribution where the red lines illustrate the theoretical prediction of Eq.\eqref{eq:sx_t}. (b)-(f) time evolution of different variables $s_x$, $s_{y,z}$, $ds_x/dt$, $ds_y/dt$ and $ds_z/dt$. (g) the distribution of field strength $B_\phi$ where the arrows denote the field direction. (h) the distribution of spin $s_r=(s_y^2+s_z^2)^{1/2}$ and the arrows mark its direction.}
\label{fig:spin_dynamics}
\end{figure}

As evident from the dependence of proton spin $s_x$ on time $t$ in Fig.~\ref{fig:spin_dynamics}b, the spin deterioration are exclusively encountered at the second stage $t>50$ fs. Accordingly, a net increment of undesired $s_y$ and $s_z$ is pronounced at $t>50$ fs (s. Fig.~\ref{fig:spin_dynamics}c), whereas a non-ignorable oscillation takes place at the first acceleration stage $t<40$ fs. Considering it is instructive to examine how the $d\mathbf{s}/dt$ governed by electric and magnetic fields, the spin differential $ds_x/dt$, $ds_y/dt$ and $ds_z/dt$ is illustrated in Fig.~\ref{fig:spin_dynamics}d-f. In non-relativistic regime $\gamma\sim1$ and $\beta\ll 1$, the cycle frequency of spin precession can be approximated as $\mathbf{\Omega}_\mathrm{non} = e(a+1)\mathbf{B}/m_pc$. At the first stage, $s_x\approx 1$ and $s_{y,z}\ll 1$ indicates that the terms incorporated with $s_{y,z}$ are negligible in relation $ds_y/dt = \Omega_zs_x-\Omega_xs_z$ and $ds_z/dt = \Omega_xs_y-\Omega_ys_x$. As a result, the accumulation of undesired spin predominantly originates from the transverse magnetic field as $ds_y/dt = e(a+1)B_zs_x/m_pc$ and $ds_z/dt = -e(a+1)B_ys_x/m_pc$. The dashed black lines in Fig.~\ref{fig:spin_dynamics}d-f correspond to the results governed by $\mathbf{\Omega}_\mathrm{non}$ under non-relativistic approximation, which are in reasonable agreement with the relativistic results. At the first stage, the oscillation of $ds_{y,z}/dt$ comes from the laser magnetic field imposed on the proton. Nevertheless, because of the periodic symmetry of the laser field, the net accumulated $s_{y,z}$ is inappreciable and thus the initial favorable spin $s_x$ characterized by $ds_x/dt = \Omega_ys_z-\Omega_zs_y$ is still largely preserved.

At the second stage, the oscillation symmetry in $ds_{y,z}/dt$ are broken (s. Fig.~\ref{fig:spin_dynamics}e-f) and a gradual increment occurs for $s_{y,z}$ (s. Fig.~\ref{fig:spin_dynamics}c), which is accompanied with the decrease of $s_x$. The reason is that a strong vortex plasma magnetic field~\cite{pukhov1996_channel,lasinski1999particle,nakamura2010_MVA_PRL}, sustained by the forward moving electron current when the laser pulse penetrates through the slab's second layer, contributes to a net accumulated precession for $s_{s,y}$. The distribution of magnetic field strength $B_{\phi}=(B_y^2+B_z^2)^{1/2}$ at $t=83$ fs averaged over $3\leqslant x[\mu m]\leqslant6$ is exhibited in Fig.~\ref{fig:spin_dynamics}g, where the field $B_{\phi}$ is along azimuthal direction and its strength is as high as 0.2 Megatesla (MT). Following the above non-relativistic assumption, we can rearrange the secondary differential of spin $s_x$ as $d^2s_x/dt^2 + \Omega_\phi^2 s_x \approx0$ and subsequently obtain the solution as 
\begin{eqnarray}
s_x(t)\approx \cos\left[\frac{|e|(a+1)B_\phi}{m_pc}(t-t_0)\right],
\label{eq:sx_t}
\end{eqnarray}
where $\Omega_\phi = e(a+1)B_\phi/m_pc$ and $t_0$ denotes the starting time of spin precession modulated by plasma vortex field $B_\phi$. The theoretically predicted $s_x(t)$ of Eq.~\eqref{eq:sx_t} is illustrated as red dashed lines in Fig.~\ref{fig:spin_dynamics}a, where $B_\phi=0.24$ MT is chosen and the prediction of $t_0=50$ fs is closest to the time-resolved spin distribution obtained from PIC simulations. This further confirms that the proton polarization loss are predominantly encountered at the second acceleration stage. The distribution of undesired transverse spin $s_r=(s_y^2+s_z^2)^{1/2}$ of protons with energy $\varepsilon_p>50$ MeV (s. Fig.~\ref{fig:spin_dynamics}h) demonstrates that the region with large $s_r$ is coincident with the strong field $B_\phi$ and $s_r$ is nearly neglectable near the central axis region $y=z=0$. The red arrows in Fig.~\ref{fig:spin_dynamics}h mark the direction of transverse spin $s_r$ and its radial outward tendency is consistent with $d\mathbf{s}_r/dt=\mathbf{\Omega}_\phi\times \mathbf{s}_x$ governed by the azimuthal plasma magnetic field $B_\phi$.

\begin{figure}[t]
\includegraphics[width=0.99\columnwidth]{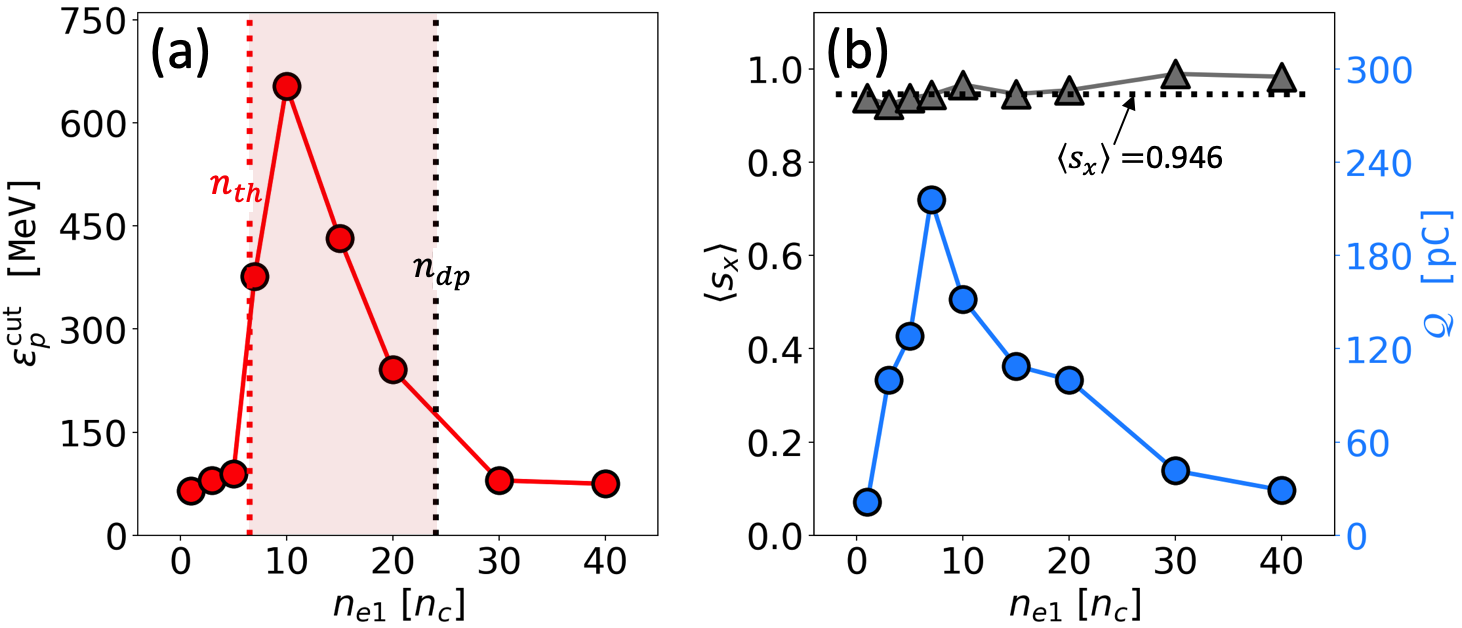}
\caption{(a) Cut-off energies of generated SPP beams versus plasma electron density $n_{e1}$ of the first layer target. (b) averaged spin polarization ratio (gray triangle) and total charge (blue circle) of the generated SPP beam as a function of density $n_{e1}$}
\label{fig:parameter_scan}
\end{figure}

To confirm the feasibility and robustness of this scheme, we examine the dependence of the acceleration efficiency on the plasma density $n_{e1}$ of the first layer target. A moderate density $n_{e1}$ is prioritized to achieve the high energy of the generated SPP beams (s. Fig~\ref{fig:parameter_scan}a). For relatively low density $n_{e1}<n_{th}$, the laser pulse readily penetrates through this transparent plasma target, where protons being not trapped but swiftly surpassed by the drift QSLEF is similar as the scenario of $n_{e1}=n_{e2}=5n_{cr}$ in Fig.~\ref{fig:schematic}d. The threshold density can be estimated via $\beta_1(n_{th})<\beta^*$, where $\beta_1(n_{th})=\mathcal{K}_+-\mathcal{K}_--1$ and $\mathcal{K}_{\pm}=[\frac{8a_0n_c}{\pi^2n_{th}}(\sqrt{1+\frac{8a_0n_c}{27\pi^2n_{th}}}\pm1)]^{1/3}$ are derived in relativistically transparent plasma~\cite{liu2020front}. Substituting $\varphi_1(\xi_u)=0.065$ and $\Tilde{p}=0$ into Eq.~\eqref{eq:threshold_beta}, one finds $\beta^*\approx0.355$ and the above density threshold can be determined as $n_{th}=6.48n_{cr}$ (s. Fig.~\ref{fig:parameter_scan}a). For relatively large density $n_{e1}>n_{dp}$, the laser pulse would be completed depleted and reflected by the accumulated overdense plasma edge before reaching the second layer due to its finite duration $\tau \sim 30$ fs. To estimate the density $n_{dp}$, we resort to the hole boring velocity $\beta_h=\sqrt{\Pi}/(1+\sqrt{\Pi})$ where $\Pi=\frac{n_{cr}}{n_{e1}}\frac{Z_im_e}{A_im_i}\frac{1}{a_0^2}$~\cite{robinson2009relativistically,schlegel2009relativistic}. The criterion 
can be interpreted as $t_1\beta_h c\sim L_1$, where $L_1=3\mu m$ the first layer thickness and $t_1$ the interaction time. By utilizing the distance relation $\beta_h t_1\approx t_1-\tau$, the upper limit density is estimated as $n_{dp}\approx 24.02n_{cr}$. Within the range of $n_{th}<n_{e1}<n_{dp}$ (s. Fig.~\ref{fig:parameter_scan}a), the proton energy is dramatically enhanced compared with the other density conditions. In addition, the averaged spin polarization ratio (s. Fig.~\ref{fig:parameter_scan}b) manifests this mechanism is favorable to preserve the proton spin polarization. $s_x=0.946$ predicted by Eq.~\eqref{eq:sx_t} indicates the insignificance of precession exerted on the protons spin by the vortex plasma magnetic field within a short time. The total charge of the generated SPP beams versus plasma density $n_{e1}$ (s. Fig.~\ref{fig:parameter_scan}b) exhibit the similar variation tendency as that of the proton cut-off energies.

In conclusion, we identified and characterized a two-stage acceleration mechanism for generation of highly energetic SPP beams. In this scenario, the protons are accelerated by the drift QSLEF twice to achieve the energy enhancement. Meanwhile the prepolarized protons substantially preserve their initial spin orientation, because the polarization loss caused by spin precession exclusively occurs in the second acceleration stage within a relatively short time. Our mechanism based on laser-plasma acceleration, realizing a SPP beam with energy near 0.5 GeV and polarization over 90\%, is an important step towards achieving the polarized ion beam quality required for the current frontiers of fundamental and nuclear physics.

\noindent\textbf{Acknowledgments}---This work has been supported by Natural Science Foundation of China (Grants No. 11921006 and No. 11535001) and National Grand Instrument Project (SQ2019YFF010006). The PIC code EPOCH was in part funded by the United Kingdom EPSRC Grants No. EP/G054950/1, No. EP/G056803/1, No. EP/G055165/1, and No. EP/M022463/1. The simulations are supported by High-performance Computing Platform of Peking University.

\section*{APPENDIX}
\begin{figure}[t]
\includegraphics[width=0.99\columnwidth]{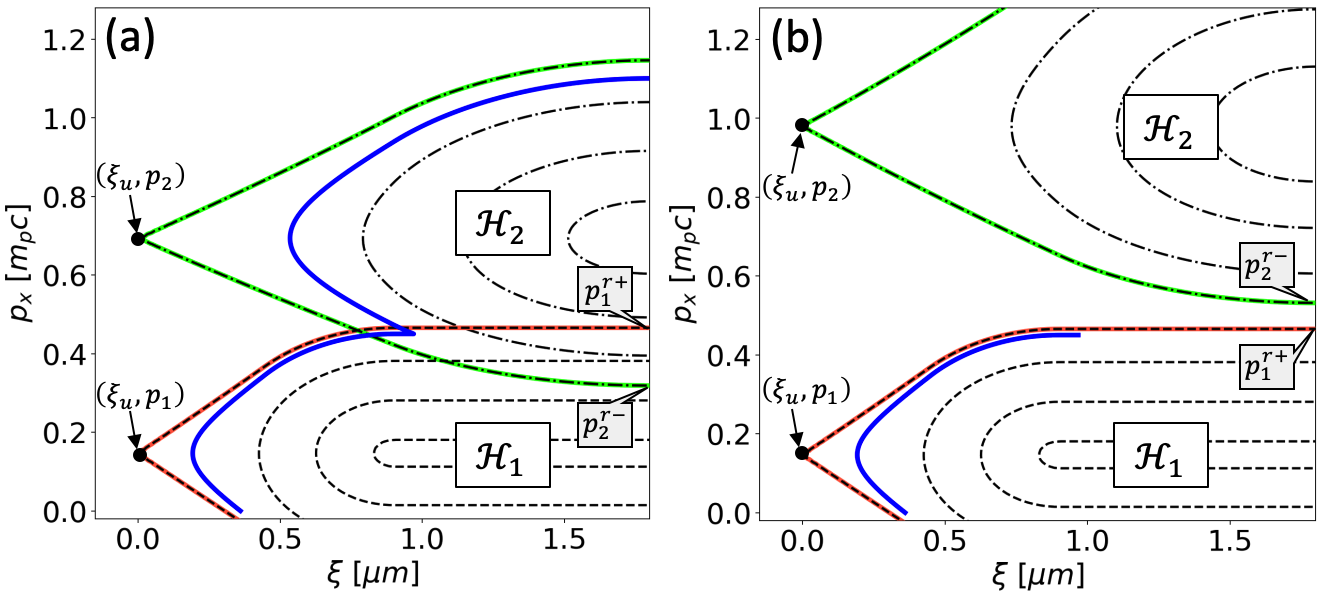}
\caption{The contours of the Hamiltonian $\mathcal{H}_1$ and $\mathcal{H}_2$ in $(\xi,p)$ space, where the blue solid lines illustrate the evolution of a representative proton. (a) for the case of $p_1^{r+}>p_2^{r-}$ while (b) for the one of $p_1^{r+}<p_2^{r-}$. The lines highlighted by red and lime color corresponds to the separatrix $\mathcal{H}_1(\xi,p)=\mathcal{H}_1(\xi_u,p_1)$ and $\mathcal{H}_2(\xi,p)=\mathcal{H}_2(\xi_u,p_2)$, respectively.}
\label{fig:hami_intersect}
\end{figure}

To explore the proton dynamics inside the two QSLEF in $(\xi,p)$ space, we adopt $\mathcal{H}_i$ and $\mathcal{H}_{i+1}$ to describe the Hamiltonian of proton evolving within the slow and fast QSLEF, respectively. In fact, the drift velocity $\beta_{i+1}$ of the fast QSLEF should not be so large that the protons reflected by the slow QSLEF is able to catch up and gain more energy. From the aspect of conserved Hamiltonian, to realize that the reflected protons from $\mathcal{H}_{i}$ can be trapped and efficiently accelerated by the $\mathcal{H}_{i+1}$, the separatrix of $\mathcal{H}_{i}$ and $\mathcal{H}_{i+1}$ should intersected with each other as shown in Fig.~\ref{fig:hami_intersect}(a). For simplicity, $i$ is set as $1$ in Fig.~\ref{fig:hami_intersect}, where the maximum [minimum] accessible momentum on the separatrix of $\mathcal{H}_1(\xi,p)=\mathcal{H}_1(\xi_u,p_1)$ [$\mathcal{H}_2(\xi,p)=\mathcal{H}_2(\xi_u,p_2)$] can be calculated as $p_1^{r+}$ [$p_2^{r-}$]. It can be found from Fig.~\ref{fig:hami_intersect}(a) that the occurrence of intersection between the two separatrices is equivalent to the condition of $p_1^{r+}>p_2^{r-}$. Therefore, the criterion for successfully coupling the proton dynamics inside these two Hamiltonian potential can be expressed as
\begin{eqnarray}\label{eqapp:criterion}
p^{r+}_{1}>p^{r-}_{2}, 
\end{eqnarray}
where 
\begin{eqnarray}
\left\{
    \begin{array}{ll}\label{eqapp:pr}
        \frac{p^{r+}_1}{m_pc}=\frac{\beta_1\mathcal{A}_1+\sqrt{\mathcal{A}_1^2+\beta_1^2-1}}{1-\beta_1^2} \\
        \frac{p^{r-}_2}{m_pc}=\frac{\beta_2\mathcal{A}_2-\sqrt{\mathcal{A}_2^2+\beta_2^2-1}}{1-\beta_2^2}
    \end{array}
\right.
\end{eqnarray}
can be found from a  previous work~\cite{gong2020proton}. Here the value of $\mathcal{A}_{1,2}=|e|\varphi_{1,2}(\xi_u)/(m_pc^2)+\sqrt{1-\beta_{1,2}^2}$ can be determined once $\beta_{1,2}$ and $\varphi_{1,2}$ are given. By employing Eq.\eqref{eqapp:pr}, the criterion of Eq.~\eqref{eqapp:criterion} can be rearranged as
\begin{widetext}
\begin{small}\begin{eqnarray} \label{th_between_two}
\frac{\beta_{1}\left[\frac{|e|\varphi_1(\xi_u)}{m_pc^2}+\sqrt{1-\beta_{1}^2}\right]+\sqrt{[\frac{|e|\varphi_1(\xi_u)}{m_pc^2}]^2+2\frac{|e|\varphi_1(\xi_u)}{m_pc^2}\sqrt{1-\beta_1^2}}}{1-\beta_1^2}>\frac{\beta_{2}\left[\frac{|e|\varphi_2(\xi_u)}{m_pc^2}+\sqrt{1-\beta_{2}^2}\right]-\sqrt{[\frac{|e|\varphi_2(\xi_u)}{m_pc^2}]^2+2\frac{|e|\varphi_2(\xi_u)}{m_pc^2}\sqrt{1-\beta_2^2}}}{1-\beta_2^2}. 
\end{eqnarray}\end{small}
\end{widetext}
Then the criterion can be analytically derived as 
\begin{widetext}
\begin{eqnarray} \label{eqapp:threshold}
\beta_2<\beta^*\equiv \frac{-\Tilde{\varphi}_2\Tilde{p}+\Tilde{p}\sqrt{1+\Tilde{p}^2}+\sqrt{\Tilde{\varphi}_2[2\sqrt{1+\Tilde{p}^2}-\Tilde{\varphi}_2]}}{1+\Tilde{p}^2},
\end{eqnarray}
\end{widetext}
where $\Tilde{p}=p^{r+}_1/(m_pc)$ and $\Tilde{\varphi}_2=|e|\varphi_2(\xi_u)/(m_pc^2)$ are adopted for convenience. It worth pointing out that the threshold of Eq.\eqref{eqapp:threshold} will return to the more general form, i.e. Eq.\eqref{eq:threshold_beta}, if the subscripts $1$ and $2$ are replaced by $i$ and $i+1$. 

The validity of criterion Eq.~\ref{eqapp:criterion} can also be illustrated by the numerically resolved proton trajectories in $(\xi,p)$ space as the solid blue lines drawn in Fig.~\ref{fig:hami_intersect}. If $p_1^{r+}>p_2^{r-}$, the proton is captured by the faster QSLEF and evolves along a contour of the $\mathcal{H}_2$ as shown in Fig.~\ref{fig:hami_intersect}(a). In contrast, if $p_1^{r+}<p_2^{r-}$, the proton does not experience an efficient acceleration at the second stage characterized by $\mathcal{H}_2$ as presented in Fig.~\ref{fig:hami_intersect}(b). For the case exhibited in Fig.~\ref{fig:hami_intersect}(a), after taking the chosen parameters $\beta_1=0.145$, $\beta_2=0.574$ and $\varphi_{1,2}(\xi_u)=0.0462$ into Eq.~\eqref{eqapp:threshold}, one can find  $\beta^*\approx0.664>\beta_2$ indicating that the accomplishment of repeated proton reflection inside these two Hamiltonian potential. By comparison, the parameters of the case in Fig.~\ref{fig:hami_intersect}(b) is same as that in Fig.~\ref{fig:hami_intersect}(a) except for $\beta_2=0.7>\beta^*$, which leads to the proton reflected in $\mathcal{H}_1$ on longer being reflected in $\mathcal{H}_2$. After performing this examination, we claim that the analytically derived threshold of Eq.~\eqref{eqapp:threshold} explicitly presents the condition of successfully coupling the two stage acceleration. It should be noted that at the limit of $\Tilde{p}\rightarrow 0$, the above threshold Eq.~\eqref{eqapp:threshold} leads to $\beta^*\equiv \sqrt{2\Tilde{\varphi}_2-\Tilde{\varphi}_2^2}$ which is the upper limit velocity to achieve efficient acceleration for initially static protons~\cite{gong2020proton}.

The realization of the two-stage coherent acceleration is determined merely by the criterion described in Eq.~\eqref{eq:threshold_beta}. The importance is to find the appropriate decreasing density combination of the double layer slabs, i.e. $n_{e1}$ and $n_{e2}$, to provide a proper drift velocity $\beta_i$ and electric potential $\varphi_i(\xi_u)$ of the QSLEF. Some plasma kinetic effects, such as plasma instabilities~\cite{Wan_2020_PRL}, may deteriorate this matching relation between the laser intensity and plasma density. Therefore, a detailed investigation and systematically simulations are needed to figure out the exact dependence of the acceleration efficiency and obtained proton energy on laser intensity.


\begin{thebibliography}{55}
\expandafter\ifx\csname natexlab\endcsname\relax\def\natexlab#1{#1}\fi
\expandafter\ifx\csname bibnamefont\endcsname\relax
  \def\bibnamefont#1{#1}\fi
\expandafter\ifx\csname bibfnamefont\endcsname\relax
  \def\bibfnamefont#1{#1}\fi
\expandafter\ifx\csname citenamefont\endcsname\relax
  \def\citenamefont#1{#1}\fi
\expandafter\ifx\csname url\endcsname\relax
  \def\url#1{\texttt{#1}}\fi
\expandafter\ifx\csname urlprefix\endcsname\relax\def\urlprefix{URL }\fi
\providecommand{\bibinfo}[2]{#2}
\providecommand{\eprint}[2][]{\url{#2}}

\bibitem[{\citenamefont{Griffiths and
  Schroeter}(2018)}]{griffiths2018introduction}
\bibinfo{author}{\bibfnamefont{D.~J.} \bibnamefont{Griffiths}}
  \bibnamefont{and} \bibinfo{author}{\bibfnamefont{D.~F.}
  \bibnamefont{Schroeter}}, \emph{\bibinfo{title}{Introduction to quantum
  mechanics}} (\bibinfo{publisher}{Cambridge University Press},
  \bibinfo{year}{2018}).

\bibitem[{\citenamefont{Mane et~al.}(2005)\citenamefont{Mane, Shatunov, and
  Yokoya}}]{mane2005_review}
\bibinfo{author}{\bibfnamefont{S.}~\bibnamefont{Mane}},
  \bibinfo{author}{\bibfnamefont{Y.~M.} \bibnamefont{Shatunov}},
  \bibnamefont{and} \bibinfo{author}{\bibfnamefont{K.}~\bibnamefont{Yokoya}},
  \bibinfo{journal}{Reports on Progress in Physics}
  \textbf{\bibinfo{volume}{68}}, \bibinfo{pages}{1997} (\bibinfo{year}{2005}).

\bibitem[{\citenamefont{Safronova et~al.}(2018)\citenamefont{Safronova, Budker,
  DeMille, Kimball, Derevianko, and Clark}}]{new_physics_from_spin}
\bibinfo{author}{\bibfnamefont{M.~S.} \bibnamefont{Safronova}},
  \bibinfo{author}{\bibfnamefont{D.}~\bibnamefont{Budker}},
  \bibinfo{author}{\bibfnamefont{D.}~\bibnamefont{DeMille}},
  \bibinfo{author}{\bibfnamefont{D.~F.~J.} \bibnamefont{Kimball}},
  \bibinfo{author}{\bibfnamefont{A.}~\bibnamefont{Derevianko}},
  \bibnamefont{and} \bibinfo{author}{\bibfnamefont{C.~W.} \bibnamefont{Clark}},
  \bibinfo{journal}{Rev. Mod. Phys.} \textbf{\bibinfo{volume}{90}},
  \bibinfo{pages}{025008} (\bibinfo{year}{2018}).

\bibitem[{\citenamefont{Aschenauer et~al.}(2019)\citenamefont{Aschenauer,
  Fazio, Lee, Mäntysaari, Page, Schenke, Ullrich, Venugopalan, and
  Zurita}}]{Rep_Pro_Phy_2019}
\bibinfo{author}{\bibfnamefont{E.~C.} \bibnamefont{Aschenauer}},
  \bibinfo{author}{\bibfnamefont{S.}~\bibnamefont{Fazio}},
  \bibinfo{author}{\bibfnamefont{J.~H.} \bibnamefont{Lee}},
  \bibinfo{author}{\bibfnamefont{H.}~\bibnamefont{Mäntysaari}},
  \bibinfo{author}{\bibfnamefont{B.~S.} \bibnamefont{Page}},
  \bibinfo{author}{\bibfnamefont{B.}~\bibnamefont{Schenke}},
  \bibinfo{author}{\bibfnamefont{T.}~\bibnamefont{Ullrich}},
  \bibinfo{author}{\bibfnamefont{R.}~\bibnamefont{Venugopalan}},
  \bibnamefont{and} \bibinfo{author}{\bibfnamefont{P.}~\bibnamefont{Zurita}},
  \bibinfo{journal}{Reports on Progress in Physics}
  \textbf{\bibinfo{volume}{82}}, \bibinfo{pages}{024301}
  (\bibinfo{year}{2019}).

\bibitem[{\citenamefont{Ji}(1997)}]{Ji_2017_PRL_nucleon_spin}
\bibinfo{author}{\bibfnamefont{X.}~\bibnamefont{Ji}}, \bibinfo{journal}{Phys.
  Rev. Lett.} \textbf{\bibinfo{volume}{78}}, \bibinfo{pages}{610}
  (\bibinfo{year}{1997}).

\bibitem[{\citenamefont{de~Florian et~al.}(2014)\citenamefont{de~Florian,
  Sassot, Stratmann, and
  Vogelsang}}]{gluon_spin_contribution_to_the_proton_spin}
\bibinfo{author}{\bibfnamefont{D.}~\bibnamefont{de~Florian}},
  \bibinfo{author}{\bibfnamefont{R.}~\bibnamefont{Sassot}},
  \bibinfo{author}{\bibfnamefont{M.}~\bibnamefont{Stratmann}},
  \bibnamefont{and}
  \bibinfo{author}{\bibfnamefont{W.}~\bibnamefont{Vogelsang}},
  \bibinfo{journal}{Phys. Rev. Lett.} \textbf{\bibinfo{volume}{113}},
  \bibinfo{pages}{012001} (\bibinfo{year}{2014}).

\bibitem[{\citenamefont{Adamczyk et~al.}(2016)\citenamefont{Adamczyk, Adkins,
  Agakishiev, Aggarwal, Ahammed, Alekseev, Aparin, Arkhipkin, Aschenauer, Attri
  et~al.}}]{investigate_QCD}
\bibinfo{author}{\bibfnamefont{L.}~\bibnamefont{Adamczyk}},
  \bibinfo{author}{\bibfnamefont{J.~K.} \bibnamefont{Adkins}},
  \bibinfo{author}{\bibfnamefont{G.}~\bibnamefont{Agakishiev}},
  \bibinfo{author}{\bibfnamefont{M.~M.} \bibnamefont{Aggarwal}},
  \bibinfo{author}{\bibfnamefont{Z.}~\bibnamefont{Ahammed}},
  \bibinfo{author}{\bibfnamefont{I.}~\bibnamefont{Alekseev}},
  \bibinfo{author}{\bibfnamefont{A.}~\bibnamefont{Aparin}},
  \bibinfo{author}{\bibfnamefont{D.}~\bibnamefont{Arkhipkin}},
  \bibinfo{author}{\bibfnamefont{E.~C.} \bibnamefont{Aschenauer}},
  \bibinfo{author}{\bibfnamefont{A.}~\bibnamefont{Attri}}, \bibnamefont{et~al.}
  (\bibinfo{collaboration}{STAR Collaboration}), \bibinfo{journal}{Phys. Rev.
  Lett.} \textbf{\bibinfo{volume}{116}}, \bibinfo{pages}{132301}
  (\bibinfo{year}{2016}).

\bibitem[{\citenamefont{Yang et~al.}(2017)\citenamefont{Yang, Sufian,
  Alexandru, Draper, Glatzmaier, Liu, and Zhao}}]{Yang_2017_QCDsim}
\bibinfo{author}{\bibfnamefont{Y.-B.} \bibnamefont{Yang}},
  \bibinfo{author}{\bibfnamefont{R.~S.} \bibnamefont{Sufian}},
  \bibinfo{author}{\bibfnamefont{A.}~\bibnamefont{Alexandru}},
  \bibinfo{author}{\bibfnamefont{T.}~\bibnamefont{Draper}},
  \bibinfo{author}{\bibfnamefont{M.~J.} \bibnamefont{Glatzmaier}},
  \bibinfo{author}{\bibfnamefont{K.-F.} \bibnamefont{Liu}}, \bibnamefont{and}
  \bibinfo{author}{\bibfnamefont{Y.}~\bibnamefont{Zhao}}
  (\bibinfo{collaboration}{\ensuremath{\chi}QCD Collaboration}),
  \bibinfo{journal}{Phys. Rev. Lett.} \textbf{\bibinfo{volume}{118}},
  \bibinfo{pages}{102001} (\bibinfo{year}{2017}).

\bibitem[{\citenamefont{Alexandrou et~al.}(2017)\citenamefont{Alexandrou,
  Constantinou, Hadjiyiannakou, Jansen, Kallidonis, Koutsou, Avilés-Casco, and
  Wiese}}]{Alexandrou2017_QCDsim}
\bibinfo{author}{\bibfnamefont{C.}~\bibnamefont{Alexandrou}},
  \bibinfo{author}{\bibfnamefont{M.}~\bibnamefont{Constantinou}},
  \bibinfo{author}{\bibfnamefont{K.}~\bibnamefont{Hadjiyiannakou}},
  \bibinfo{author}{\bibfnamefont{K.}~\bibnamefont{Jansen}},
  \bibinfo{author}{\bibfnamefont{C.}~\bibnamefont{Kallidonis}},
  \bibinfo{author}{\bibfnamefont{G.}~\bibnamefont{Koutsou}},
  \bibinfo{author}{\bibfnamefont{A.~V.} \bibnamefont{Avilés-Casco}},
  \bibnamefont{and} \bibinfo{author}{\bibfnamefont{C.}~\bibnamefont{Wiese}},
  \bibinfo{journal}{Physical Review Letters} \textbf{\bibinfo{volume}{119}},
  \bibinfo{pages}{142002} (\bibinfo{year}{2017}).

\bibitem[{\citenamefont{Adamczyk et~al.}(2014)\citenamefont{Adamczyk, Adkins,
  Agakishiev, Aggarwal, Ahammed, Alekseev, Alford, Anson, Aparin, Arkhipkin
  et~al.}}]{parity_violating_spin_asymmetry_PRL}
\bibinfo{author}{\bibfnamefont{L.}~\bibnamefont{Adamczyk}},
  \bibinfo{author}{\bibfnamefont{J.~K.} \bibnamefont{Adkins}},
  \bibinfo{author}{\bibfnamefont{G.}~\bibnamefont{Agakishiev}},
  \bibinfo{author}{\bibfnamefont{M.~M.} \bibnamefont{Aggarwal}},
  \bibinfo{author}{\bibfnamefont{Z.}~\bibnamefont{Ahammed}},
  \bibinfo{author}{\bibfnamefont{I.}~\bibnamefont{Alekseev}},
  \bibinfo{author}{\bibfnamefont{J.}~\bibnamefont{Alford}},
  \bibinfo{author}{\bibfnamefont{C.~D.} \bibnamefont{Anson}},
  \bibinfo{author}{\bibfnamefont{A.}~\bibnamefont{Aparin}},
  \bibinfo{author}{\bibfnamefont{D.}~\bibnamefont{Arkhipkin}},
  \bibnamefont{et~al.} (\bibinfo{collaboration}{STAR Collaboration}),
  \bibinfo{journal}{Phys. Rev. Lett.} \textbf{\bibinfo{volume}{113}},
  \bibinfo{pages}{072301} (\bibinfo{year}{2014}).

\bibitem[{\citenamefont{Jaeckel et~al.}(2020)\citenamefont{Jaeckel, Lamont, and
  Vallée}}]{np_2020_beyond_standard}
\bibinfo{author}{\bibfnamefont{J.}~\bibnamefont{Jaeckel}},
  \bibinfo{author}{\bibfnamefont{M.}~\bibnamefont{Lamont}}, \bibnamefont{and}
  \bibinfo{author}{\bibfnamefont{C.}~\bibnamefont{Vallée}},
  \bibinfo{journal}{Nature Physics} \textbf{\bibinfo{volume}{16}},
  \bibinfo{pages}{393} (\bibinfo{year}{2020}).

\bibitem[{\citenamefont{Rosen}(1967)}]{rosen1967polarized_Science}
\bibinfo{author}{\bibfnamefont{L.}~\bibnamefont{Rosen}},
  \bibinfo{journal}{Science} \textbf{\bibinfo{volume}{157}},
  \bibinfo{pages}{1127} (\bibinfo{year}{1967}).

\bibitem[{\citenamefont{Tojo et~al.}(2002)\citenamefont{Tojo, Alekseev, Bai,
  Bassalleck, Bunce, Deshpande, Doskow, Eilerts, Fields, Goto
  et~al.}}]{Tojo_PRL_2002}
\bibinfo{author}{\bibfnamefont{J.}~\bibnamefont{Tojo}},
  \bibinfo{author}{\bibfnamefont{I.}~\bibnamefont{Alekseev}},
  \bibinfo{author}{\bibfnamefont{M.}~\bibnamefont{Bai}},
  \bibinfo{author}{\bibfnamefont{B.}~\bibnamefont{Bassalleck}},
  \bibinfo{author}{\bibfnamefont{G.}~\bibnamefont{Bunce}},
  \bibinfo{author}{\bibfnamefont{A.}~\bibnamefont{Deshpande}},
  \bibinfo{author}{\bibfnamefont{J.}~\bibnamefont{Doskow}},
  \bibinfo{author}{\bibfnamefont{S.}~\bibnamefont{Eilerts}},
  \bibinfo{author}{\bibfnamefont{D.~E.} \bibnamefont{Fields}},
  \bibinfo{author}{\bibfnamefont{Y.}~\bibnamefont{Goto}}, \bibnamefont{et~al.},
  \bibinfo{journal}{Phys. Rev. Lett.} \textbf{\bibinfo{volume}{89}},
  \bibinfo{pages}{052302} (\bibinfo{year}{2002}).

\bibitem[{\citenamefont{Allgower et~al.}(2002)\citenamefont{Allgower, Krueger,
  Kasprzyk, Spinka, Underwood, Yokosawa, Bunce, Huang, Makdisi, Roser
  et~al.}}]{Allgower_PRD_2002}
\bibinfo{author}{\bibfnamefont{C.~E.} \bibnamefont{Allgower}},
  \bibinfo{author}{\bibfnamefont{K.~W.} \bibnamefont{Krueger}},
  \bibinfo{author}{\bibfnamefont{T.~E.} \bibnamefont{Kasprzyk}},
  \bibinfo{author}{\bibfnamefont{H.~M.} \bibnamefont{Spinka}},
  \bibinfo{author}{\bibfnamefont{D.~G.} \bibnamefont{Underwood}},
  \bibinfo{author}{\bibfnamefont{A.}~\bibnamefont{Yokosawa}},
  \bibinfo{author}{\bibfnamefont{G.}~\bibnamefont{Bunce}},
  \bibinfo{author}{\bibfnamefont{H.}~\bibnamefont{Huang}},
  \bibinfo{author}{\bibfnamefont{Y.}~\bibnamefont{Makdisi}},
  \bibinfo{author}{\bibfnamefont{T.}~\bibnamefont{Roser}}, \bibnamefont{et~al.}
  (\bibinfo{collaboration}{E925 Collaboration}), \bibinfo{journal}{Phys. Rev.
  D} \textbf{\bibinfo{volume}{65}}, \bibinfo{pages}{092008}
  (\bibinfo{year}{2002}).

\bibitem[{\citenamefont{Glavish et~al.}(1972)\citenamefont{Glavish, Hanna,
  Avida, Boyd, Chang, and Diener}}]{glavish1972giant_GDR}
\bibinfo{author}{\bibfnamefont{H.}~\bibnamefont{Glavish}},
  \bibinfo{author}{\bibfnamefont{S.}~\bibnamefont{Hanna}},
  \bibinfo{author}{\bibfnamefont{R.}~\bibnamefont{Avida}},
  \bibinfo{author}{\bibfnamefont{R.}~\bibnamefont{Boyd}},
  \bibinfo{author}{\bibfnamefont{C.}~\bibnamefont{Chang}}, \bibnamefont{and}
  \bibinfo{author}{\bibfnamefont{E.}~\bibnamefont{Diener}},
  \bibinfo{journal}{Physical Review Letters} \textbf{\bibinfo{volume}{28}},
  \bibinfo{pages}{766} (\bibinfo{year}{1972}).

\bibitem[{\citenamefont{Kitching et~al.}(1986)\citenamefont{Kitching, Hutcheon,
  Michaelian, Abegg, Coombes, Dawson, Fielding, Gaillard, Green, Greeniaus
  et~al.}}]{kitching1986polarized_pp_bremsstrahlung}
\bibinfo{author}{\bibfnamefont{P.}~\bibnamefont{Kitching}},
  \bibinfo{author}{\bibfnamefont{D.}~\bibnamefont{Hutcheon}},
  \bibinfo{author}{\bibfnamefont{K.}~\bibnamefont{Michaelian}},
  \bibinfo{author}{\bibfnamefont{R.}~\bibnamefont{Abegg}},
  \bibinfo{author}{\bibfnamefont{G.}~\bibnamefont{Coombes}},
  \bibinfo{author}{\bibfnamefont{W.}~\bibnamefont{Dawson}},
  \bibinfo{author}{\bibfnamefont{H.}~\bibnamefont{Fielding}},
  \bibinfo{author}{\bibfnamefont{G.}~\bibnamefont{Gaillard}},
  \bibinfo{author}{\bibfnamefont{P.}~\bibnamefont{Green}},
  \bibinfo{author}{\bibfnamefont{L.}~\bibnamefont{Greeniaus}},
  \bibnamefont{et~al.}, \bibinfo{journal}{Physical review letters}
  \textbf{\bibinfo{volume}{57}}, \bibinfo{pages}{2363} (\bibinfo{year}{1986}).

\bibitem[{\citenamefont{Kee et~al.}(2013)\citenamefont{Kee, Zhu, Hildenbrand,
  V{\o}llestad, Sanders, and O’Hayre}}]{kee2013modeling}
\bibinfo{author}{\bibfnamefont{R.~J.} \bibnamefont{Kee}},
  \bibinfo{author}{\bibfnamefont{H.}~\bibnamefont{Zhu}},
  \bibinfo{author}{\bibfnamefont{B.~W.} \bibnamefont{Hildenbrand}},
  \bibinfo{author}{\bibfnamefont{E.}~\bibnamefont{V{\o}llestad}},
  \bibinfo{author}{\bibfnamefont{M.~D.} \bibnamefont{Sanders}},
  \bibnamefont{and} \bibinfo{author}{\bibfnamefont{R.~P.}
  \bibnamefont{O’Hayre}}, \bibinfo{journal}{Journal of The Electrochemical
  Society} \textbf{\bibinfo{volume}{160}}, \bibinfo{pages}{F290}
  (\bibinfo{year}{2013}).

\bibitem[{\citenamefont{Zimmer et~al.}(2016)\citenamefont{Zimmer, Jouve, and
  Stuhrmann}}]{zimmer2016polarized}
\bibinfo{author}{\bibfnamefont{O.}~\bibnamefont{Zimmer}},
  \bibinfo{author}{\bibfnamefont{H.~M.} \bibnamefont{Jouve}}, \bibnamefont{and}
  \bibinfo{author}{\bibfnamefont{H.~B.} \bibnamefont{Stuhrmann}},
  \bibinfo{journal}{IUCrJ} \textbf{\bibinfo{volume}{3}}, \bibinfo{pages}{326}
  (\bibinfo{year}{2016}).

\bibitem[{\citenamefont{Bai et~al.}(2006)\citenamefont{Bai, Roser, Ahrens,
  Alekseev, Alessi, Beebe-Wang, Blaskiewicz, Bravar, Brennan, Bruno
  et~al.}}]{bai2006polarized}
\bibinfo{author}{\bibfnamefont{M.}~\bibnamefont{Bai}},
  \bibinfo{author}{\bibfnamefont{T.}~\bibnamefont{Roser}},
  \bibinfo{author}{\bibfnamefont{L.}~\bibnamefont{Ahrens}},
  \bibinfo{author}{\bibfnamefont{I.}~\bibnamefont{Alekseev}},
  \bibinfo{author}{\bibfnamefont{J.}~\bibnamefont{Alessi}},
  \bibinfo{author}{\bibfnamefont{J.}~\bibnamefont{Beebe-Wang}},
  \bibinfo{author}{\bibfnamefont{M.}~\bibnamefont{Blaskiewicz}},
  \bibinfo{author}{\bibfnamefont{A.}~\bibnamefont{Bravar}},
  \bibinfo{author}{\bibfnamefont{J.}~\bibnamefont{Brennan}},
  \bibinfo{author}{\bibfnamefont{D.}~\bibnamefont{Bruno}},
  \bibnamefont{et~al.}, \bibinfo{journal}{Physical review letters}
  \textbf{\bibinfo{volume}{96}}, \bibinfo{pages}{174801}
  (\bibinfo{year}{2006}).

\bibitem[{\citenamefont{Derbenev and
  Kondratenko}(1975)}]{derbenev1975acceleration_sebrian_snake}
\bibinfo{author}{\bibfnamefont{Y.~S.} \bibnamefont{Derbenev}} \bibnamefont{and}
  \bibinfo{author}{\bibfnamefont{A.~M.} \bibnamefont{Kondratenko}},
  \bibinfo{journal}{Doklady Akademii Nauk SSSR} \textbf{\bibinfo{volume}{223}},
  \bibinfo{pages}{830} (\bibinfo{year}{1975}).

\bibitem[{\citenamefont{Strickland and Mourou}(1985)}]{CPA_1985compression}
\bibinfo{author}{\bibfnamefont{D.}~\bibnamefont{Strickland}} \bibnamefont{and}
  \bibinfo{author}{\bibfnamefont{G.}~\bibnamefont{Mourou}},
  \bibinfo{journal}{Opt. Commun.} \textbf{\bibinfo{volume}{55}},
  \bibinfo{pages}{447} (\bibinfo{year}{1985}).

\bibitem[{\citenamefont{Mourou et~al.}(2006)\citenamefont{Mourou, Tajima, and
  Bulanov}}]{Mourou_etal_2006}
\bibinfo{author}{\bibfnamefont{G.~A.} \bibnamefont{Mourou}},
  \bibinfo{author}{\bibfnamefont{T.}~\bibnamefont{Tajima}}, \bibnamefont{and}
  \bibinfo{author}{\bibfnamefont{S.~V.} \bibnamefont{Bulanov}},
  \bibinfo{journal}{Rev. Mod. Phys.} \textbf{\bibinfo{volume}{78}},
  \bibinfo{pages}{309} (\bibinfo{year}{2006}).

\bibitem[{\citenamefont{Danson et~al.}(2019)\citenamefont{Danson, Haefner,
  Bromage, Butcher, Chanteloup, Chowdhury, Galvanauskas, Gizzi, Hein, Hillier
  et~al.}}]{danson2019petawatt_laser}
\bibinfo{author}{\bibfnamefont{C.~N.} \bibnamefont{Danson}},
  \bibinfo{author}{\bibfnamefont{C.}~\bibnamefont{Haefner}},
  \bibinfo{author}{\bibfnamefont{J.}~\bibnamefont{Bromage}},
  \bibinfo{author}{\bibfnamefont{T.}~\bibnamefont{Butcher}},
  \bibinfo{author}{\bibfnamefont{J.-C.~F.} \bibnamefont{Chanteloup}},
  \bibinfo{author}{\bibfnamefont{E.~A.} \bibnamefont{Chowdhury}},
  \bibinfo{author}{\bibfnamefont{A.}~\bibnamefont{Galvanauskas}},
  \bibinfo{author}{\bibfnamefont{L.~A.} \bibnamefont{Gizzi}},
  \bibinfo{author}{\bibfnamefont{J.}~\bibnamefont{Hein}},
  \bibinfo{author}{\bibfnamefont{D.~I.} \bibnamefont{Hillier}},
  \bibnamefont{et~al.}, \bibinfo{journal}{High Power Laser Science and
  Engineering} \textbf{\bibinfo{volume}{7}} (\bibinfo{year}{2019}).

\bibitem[{\citenamefont{Tajima and Dawson}(1979)}]{tajima1979}
\bibinfo{author}{\bibfnamefont{T.}~\bibnamefont{Tajima}} \bibnamefont{and}
  \bibinfo{author}{\bibfnamefont{J.}~\bibnamefont{Dawson}},
  \bibinfo{journal}{Physical Review Letters} \textbf{\bibinfo{volume}{43}},
  \bibinfo{pages}{267} (\bibinfo{year}{1979}).

\bibitem[{\citenamefont{Esarey et~al.}(2009)\citenamefont{Esarey, Schroeder,
  and Leemans}}]{Esarey_2009_RMP}
\bibinfo{author}{\bibfnamefont{E.}~\bibnamefont{Esarey}},
  \bibinfo{author}{\bibfnamefont{C.~B.} \bibnamefont{Schroeder}},
  \bibnamefont{and} \bibinfo{author}{\bibfnamefont{W.~P.}
  \bibnamefont{Leemans}}, \bibinfo{journal}{Rev. Mod. Phys.}
  \textbf{\bibinfo{volume}{81}}, \bibinfo{pages}{1229} (\bibinfo{year}{2009}).

\bibitem[{\citenamefont{Macchi et~al.}(2013)\citenamefont{Macchi, Borghesi, and
  Passoni}}]{macchi_2013_RMP}
\bibinfo{author}{\bibfnamefont{A.}~\bibnamefont{Macchi}},
  \bibinfo{author}{\bibfnamefont{M.}~\bibnamefont{Borghesi}}, \bibnamefont{and}
  \bibinfo{author}{\bibfnamefont{M.}~\bibnamefont{Passoni}},
  \bibinfo{journal}{Reviews of Modern Physics} \textbf{\bibinfo{volume}{85}},
  \bibinfo{pages}{751} (\bibinfo{year}{2013}).

\bibitem[{\citenamefont{Rakitzis et~al.}(2003)\citenamefont{Rakitzis,
  Samartzis, Toomes, Kitsopoulos, Brown, Balint-Kurti, Vasyutinskii, and
  Beswick}}]{rakitzis2003spin_science}
\bibinfo{author}{\bibfnamefont{T.}~\bibnamefont{Rakitzis}},
  \bibinfo{author}{\bibfnamefont{P.}~\bibnamefont{Samartzis}},
  \bibinfo{author}{\bibfnamefont{R.}~\bibnamefont{Toomes}},
  \bibinfo{author}{\bibfnamefont{T.}~\bibnamefont{Kitsopoulos}},
  \bibinfo{author}{\bibfnamefont{A.}~\bibnamefont{Brown}},
  \bibinfo{author}{\bibfnamefont{G.}~\bibnamefont{Balint-Kurti}},
  \bibinfo{author}{\bibfnamefont{O.}~\bibnamefont{Vasyutinskii}},
  \bibnamefont{and} \bibinfo{author}{\bibfnamefont{J.}~\bibnamefont{Beswick}},
  \bibinfo{journal}{Science} \textbf{\bibinfo{volume}{300}},
  \bibinfo{pages}{1936} (\bibinfo{year}{2003}).

\bibitem[{\citenamefont{Sofikitis et~al.}(2008)\citenamefont{Sofikitis,
  Rubio-Lago, Bougas, Alexander, and Rakitzis}}]{Sofikitis2008Laser_JCP}
\bibinfo{author}{\bibfnamefont{D.}~\bibnamefont{Sofikitis}},
  \bibinfo{author}{\bibfnamefont{L.}~\bibnamefont{Rubio-Lago}},
  \bibinfo{author}{\bibfnamefont{L.}~\bibnamefont{Bougas}},
  \bibinfo{author}{\bibfnamefont{A.~J.} \bibnamefont{Alexander}},
  \bibnamefont{and} \bibinfo{author}{\bibfnamefont{T.~P.}
  \bibnamefont{Rakitzis}}, \bibinfo{journal}{Journal of Chemical Physics}
  \textbf{\bibinfo{volume}{129}}, \bibinfo{pages}{144302}
  (\bibinfo{year}{2008}).

\bibitem[{\citenamefont{Sofikitis et~al.}(2017)\citenamefont{Sofikitis, Glodic,
  Koumarianou, Jiang, Bougas, Samartzis, Andreev, and
  Rakitzis}}]{sofikitis2017spin_PRL}
\bibinfo{author}{\bibfnamefont{D.}~\bibnamefont{Sofikitis}},
  \bibinfo{author}{\bibfnamefont{P.}~\bibnamefont{Glodic}},
  \bibinfo{author}{\bibfnamefont{G.}~\bibnamefont{Koumarianou}},
  \bibinfo{author}{\bibfnamefont{H.}~\bibnamefont{Jiang}},
  \bibinfo{author}{\bibfnamefont{L.}~\bibnamefont{Bougas}},
  \bibinfo{author}{\bibfnamefont{P.~C.} \bibnamefont{Samartzis}},
  \bibinfo{author}{\bibfnamefont{A.}~\bibnamefont{Andreev}}, \bibnamefont{and}
  \bibinfo{author}{\bibfnamefont{T.~P.} \bibnamefont{Rakitzis}},
  \bibinfo{journal}{Phys. Rev. Lett.} \textbf{\bibinfo{volume}{118}},
  \bibinfo{pages}{233401} (\bibinfo{year}{2017}).

\bibitem[{\citenamefont{Boulogiannis et~al.}(2019)\citenamefont{Boulogiannis,
  Kannis, Katsoprinakis, Sofikitis, and Rakitzis}}]{boulogiannis2019spin}
\bibinfo{author}{\bibfnamefont{G.~K.} \bibnamefont{Boulogiannis}},
  \bibinfo{author}{\bibfnamefont{C.~S.} \bibnamefont{Kannis}},
  \bibinfo{author}{\bibfnamefont{G.~E.} \bibnamefont{Katsoprinakis}},
  \bibinfo{author}{\bibfnamefont{D.}~\bibnamefont{Sofikitis}},
  \bibnamefont{and} \bibinfo{author}{\bibfnamefont{T.~P.}
  \bibnamefont{Rakitzis}}, \bibinfo{journal}{The Journal of Physical Chemistry
  A} \textbf{\bibinfo{volume}{123}}, \bibinfo{pages}{8130}
  (\bibinfo{year}{2019}).

\bibitem[{\citenamefont{Sofikitis et~al.}(2018)\citenamefont{Sofikitis, Kannis,
  Boulogiannis, and Rakitzis}}]{sofikitis2018spin_PRL}
\bibinfo{author}{\bibfnamefont{D.}~\bibnamefont{Sofikitis}},
  \bibinfo{author}{\bibfnamefont{C.~S.} \bibnamefont{Kannis}},
  \bibinfo{author}{\bibfnamefont{G.~K.} \bibnamefont{Boulogiannis}},
  \bibnamefont{and} \bibinfo{author}{\bibfnamefont{T.~P.}
  \bibnamefont{Rakitzis}}, \bibinfo{journal}{Phys. Rev. Lett.}
  \textbf{\bibinfo{volume}{121}}, \bibinfo{pages}{083001}
  (\bibinfo{year}{2018}).

\bibitem[{\citenamefont{Wen et~al.}(2019)\citenamefont{Wen, Tamburini, and
  Keitel}}]{wen2019polarized}
\bibinfo{author}{\bibfnamefont{M.}~\bibnamefont{Wen}},
  \bibinfo{author}{\bibfnamefont{M.}~\bibnamefont{Tamburini}},
  \bibnamefont{and} \bibinfo{author}{\bibfnamefont{C.~H.}
  \bibnamefont{Keitel}}, \bibinfo{journal}{Physical review letters}
  \textbf{\bibinfo{volume}{122}}, \bibinfo{pages}{214801}
  (\bibinfo{year}{2019}).

\bibitem[{\citenamefont{Wu et~al.}(2019{\natexlab{a}})\citenamefont{Wu, Ji,
  Geng, Yu, Wang, Feng, Guo, Wang, Qin, Yan et~al.}}]{wu2019polarized_NJP}
\bibinfo{author}{\bibfnamefont{Y.}~\bibnamefont{Wu}},
  \bibinfo{author}{\bibfnamefont{L.}~\bibnamefont{Ji}},
  \bibinfo{author}{\bibfnamefont{X.}~\bibnamefont{Geng}},
  \bibinfo{author}{\bibfnamefont{Q.}~\bibnamefont{Yu}},
  \bibinfo{author}{\bibfnamefont{N.}~\bibnamefont{Wang}},
  \bibinfo{author}{\bibfnamefont{B.}~\bibnamefont{Feng}},
  \bibinfo{author}{\bibfnamefont{Z.}~\bibnamefont{Guo}},
  \bibinfo{author}{\bibfnamefont{W.}~\bibnamefont{Wang}},
  \bibinfo{author}{\bibfnamefont{C.}~\bibnamefont{Qin}},
  \bibinfo{author}{\bibfnamefont{X.}~\bibnamefont{Yan}}, \bibnamefont{et~al.},
  \bibinfo{journal}{New Journal of Physics} \textbf{\bibinfo{volume}{21}},
  \bibinfo{pages}{073052} (\bibinfo{year}{2019}{\natexlab{a}}).

\bibitem[{\citenamefont{Wu et~al.}(2019{\natexlab{b}})\citenamefont{Wu, Ji,
  Geng, Yu, Wang, Feng, Guo, Wang, Qin, Yan et~al.}}]{wu2019polarized_PRE}
\bibinfo{author}{\bibfnamefont{Y.}~\bibnamefont{Wu}},
  \bibinfo{author}{\bibfnamefont{L.}~\bibnamefont{Ji}},
  \bibinfo{author}{\bibfnamefont{X.}~\bibnamefont{Geng}},
  \bibinfo{author}{\bibfnamefont{Q.}~\bibnamefont{Yu}},
  \bibinfo{author}{\bibfnamefont{N.}~\bibnamefont{Wang}},
  \bibinfo{author}{\bibfnamefont{B.}~\bibnamefont{Feng}},
  \bibinfo{author}{\bibfnamefont{Z.}~\bibnamefont{Guo}},
  \bibinfo{author}{\bibfnamefont{W.}~\bibnamefont{Wang}},
  \bibinfo{author}{\bibfnamefont{C.}~\bibnamefont{Qin}},
  \bibinfo{author}{\bibfnamefont{X.}~\bibnamefont{Yan}}, \bibnamefont{et~al.},
  \bibinfo{journal}{Physical Review E} \textbf{\bibinfo{volume}{100}},
  \bibinfo{pages}{043202} (\bibinfo{year}{2019}{\natexlab{b}}).

\bibitem[{\citenamefont{Hu et~al.}(2020)\citenamefont{Hu, Zhou, Tao, Lv, Zou,
  and Ding}}]{Hu_2020_PRE}
\bibinfo{author}{\bibfnamefont{R.}~\bibnamefont{Hu}},
  \bibinfo{author}{\bibfnamefont{H.}~\bibnamefont{Zhou}},
  \bibinfo{author}{\bibfnamefont{Z.}~\bibnamefont{Tao}},
  \bibinfo{author}{\bibfnamefont{M.}~\bibnamefont{Lv}},
  \bibinfo{author}{\bibfnamefont{S.}~\bibnamefont{Zou}}, \bibnamefont{and}
  \bibinfo{author}{\bibfnamefont{Y.}~\bibnamefont{Ding}},
  \bibinfo{journal}{Phys. Rev. E} \textbf{\bibinfo{volume}{102}},
  \bibinfo{pages}{043215} (\bibinfo{year}{2020}).

\bibitem[{\citenamefont{Hützen et~al.}(2019)\citenamefont{Hützen, Thomas,
  Böker, Engels, Gebel, Lehrach, Pukhov, Rakitzis, Sofikitis, and
  Büscher}}]{H2019Polarized}
\bibinfo{author}{\bibfnamefont{A.}~\bibnamefont{Hützen}},
  \bibinfo{author}{\bibfnamefont{J.}~\bibnamefont{Thomas}},
  \bibinfo{author}{\bibfnamefont{J.}~\bibnamefont{Böker}},
  \bibinfo{author}{\bibfnamefont{R.}~\bibnamefont{Engels}},
  \bibinfo{author}{\bibfnamefont{R.}~\bibnamefont{Gebel}},
  \bibinfo{author}{\bibfnamefont{A.}~\bibnamefont{Lehrach}},
  \bibinfo{author}{\bibfnamefont{A.}~\bibnamefont{Pukhov}},
  \bibinfo{author}{\bibfnamefont{T.~P.} \bibnamefont{Rakitzis}},
  \bibinfo{author}{\bibfnamefont{D.}~\bibnamefont{Sofikitis}},
  \bibnamefont{and} \bibinfo{author}{\bibfnamefont{M.}~\bibnamefont{Büscher}},
  \bibinfo{journal}{High Power Laser Science and Engineering}
  \textbf{\bibinfo{volume}{7}} (\bibinfo{year}{2019}).

\bibitem[{\citenamefont{Büscher et~al.}(2019)\citenamefont{Büscher, Hützen,
  Engin, Thomas, Pukhov, Böker, Gebel, Lehrach, Engels, Peter~Rakitzis
  et~al.}}]{Buscher_2019}
\bibinfo{author}{\bibfnamefont{M.}~\bibnamefont{Büscher}},
  \bibinfo{author}{\bibfnamefont{A.}~\bibnamefont{Hützen}},
  \bibinfo{author}{\bibfnamefont{I.}~\bibnamefont{Engin}},
  \bibinfo{author}{\bibfnamefont{J.}~\bibnamefont{Thomas}},
  \bibinfo{author}{\bibfnamefont{A.}~\bibnamefont{Pukhov}},
  \bibinfo{author}{\bibfnamefont{J.}~\bibnamefont{Böker}},
  \bibinfo{author}{\bibfnamefont{R.}~\bibnamefont{Gebel}},
  \bibinfo{author}{\bibfnamefont{A.}~\bibnamefont{Lehrach}},
  \bibinfo{author}{\bibfnamefont{R.}~\bibnamefont{Engels}},
  \bibinfo{author}{\bibfnamefont{T.}~\bibnamefont{Peter~Rakitzis}},
  \bibnamefont{et~al.}, \bibinfo{journal}{International Journal of Modern
  Physics A} \textbf{\bibinfo{volume}{34}}, \bibinfo{pages}{1942028}
  (\bibinfo{year}{2019}).

\bibitem[{\citenamefont{Jin et~al.}(2020)\citenamefont{Jin, Wen, Zhang,
  H\"utzen, Thomas, B\"uscher, and Shen}}]{jin2020spin}
\bibinfo{author}{\bibfnamefont{L.}~\bibnamefont{Jin}},
  \bibinfo{author}{\bibfnamefont{M.}~\bibnamefont{Wen}},
  \bibinfo{author}{\bibfnamefont{X.}~\bibnamefont{Zhang}},
  \bibinfo{author}{\bibfnamefont{A.}~\bibnamefont{H\"utzen}},
  \bibinfo{author}{\bibfnamefont{J.}~\bibnamefont{Thomas}},
  \bibinfo{author}{\bibfnamefont{M.}~\bibnamefont{B\"uscher}},
  \bibnamefont{and} \bibinfo{author}{\bibfnamefont{B.}~\bibnamefont{Shen}},
  \bibinfo{journal}{Phys. Rev. E} \textbf{\bibinfo{volume}{102}},
  \bibinfo{pages}{011201} (\bibinfo{year}{2020}).

\bibitem[{\citenamefont{Arber et~al.}(2015)\citenamefont{Arber, Bennett, Brady,
  Lawrence-Douglas, Ramsay, Sircombe, Gillies, Evans, Schmitz, Bell
  et~al.}}]{arber2015contemporary}
\bibinfo{author}{\bibfnamefont{T.}~\bibnamefont{Arber}},
  \bibinfo{author}{\bibfnamefont{K.}~\bibnamefont{Bennett}},
  \bibinfo{author}{\bibfnamefont{C.}~\bibnamefont{Brady}},
  \bibinfo{author}{\bibfnamefont{A.}~\bibnamefont{Lawrence-Douglas}},
  \bibinfo{author}{\bibfnamefont{M.}~\bibnamefont{Ramsay}},
  \bibinfo{author}{\bibfnamefont{N.}~\bibnamefont{Sircombe}},
  \bibinfo{author}{\bibfnamefont{P.}~\bibnamefont{Gillies}},
  \bibinfo{author}{\bibfnamefont{R.}~\bibnamefont{Evans}},
  \bibinfo{author}{\bibfnamefont{H.}~\bibnamefont{Schmitz}},
  \bibinfo{author}{\bibfnamefont{A.}~\bibnamefont{Bell}}, \bibnamefont{et~al.},
  \bibinfo{journal}{Plasma Physics and Controlled Fusion}
  \textbf{\bibinfo{volume}{57}}, \bibinfo{pages}{113001}
  (\bibinfo{year}{2015}).

\bibitem[{\citenamefont{Ma et~al.}(2007)\citenamefont{Ma, Song, Yang, Zhang,
  Zhao, Sun, Ren, Liu, Liu, Shen et~al.}}]{ma2007directly}
\bibinfo{author}{\bibfnamefont{W.}~\bibnamefont{Ma}},
  \bibinfo{author}{\bibfnamefont{L.}~\bibnamefont{Song}},
  \bibinfo{author}{\bibfnamefont{R.}~\bibnamefont{Yang}},
  \bibinfo{author}{\bibfnamefont{T.}~\bibnamefont{Zhang}},
  \bibinfo{author}{\bibfnamefont{Y.}~\bibnamefont{Zhao}},
  \bibinfo{author}{\bibfnamefont{L.}~\bibnamefont{Sun}},
  \bibinfo{author}{\bibfnamefont{Y.}~\bibnamefont{Ren}},
  \bibinfo{author}{\bibfnamefont{D.}~\bibnamefont{Liu}},
  \bibinfo{author}{\bibfnamefont{L.}~\bibnamefont{Liu}},
  \bibinfo{author}{\bibfnamefont{J.}~\bibnamefont{Shen}}, \bibnamefont{et~al.},
  \bibinfo{journal}{Nano Letters} \textbf{\bibinfo{volume}{7}},
  \bibinfo{pages}{2307} (\bibinfo{year}{2007}).

\bibitem[{\citenamefont{Zhang et~al.}(2007)\citenamefont{Zhang, Shen, Li, Jin,
  and Wang}}]{zhang2007multistaged}
\bibinfo{author}{\bibfnamefont{X.}~\bibnamefont{Zhang}},
  \bibinfo{author}{\bibfnamefont{B.}~\bibnamefont{Shen}},
  \bibinfo{author}{\bibfnamefont{X.}~\bibnamefont{Li}},
  \bibinfo{author}{\bibfnamefont{Z.}~\bibnamefont{Jin}}, \bibnamefont{and}
  \bibinfo{author}{\bibfnamefont{F.}~\bibnamefont{Wang}},
  \bibinfo{journal}{Physics of Plasmas} \textbf{\bibinfo{volume}{14}},
  \bibinfo{pages}{073101} (\bibinfo{year}{2007}).

\bibitem[{\citenamefont{Robinson et~al.}(2008)\citenamefont{Robinson, Zepf,
  Kar, Evans, and Bellei}}]{robinson2008radiation}
\bibinfo{author}{\bibfnamefont{A.}~\bibnamefont{Robinson}},
  \bibinfo{author}{\bibfnamefont{M.}~\bibnamefont{Zepf}},
  \bibinfo{author}{\bibfnamefont{S.}~\bibnamefont{Kar}},
  \bibinfo{author}{\bibfnamefont{R.}~\bibnamefont{Evans}}, \bibnamefont{and}
  \bibinfo{author}{\bibfnamefont{C.}~\bibnamefont{Bellei}},
  \bibinfo{journal}{New journal of Physics} \textbf{\bibinfo{volume}{10}},
  \bibinfo{pages}{013021} (\bibinfo{year}{2008}).

\bibitem[{\citenamefont{Robinson et~al.}(2009)\citenamefont{Robinson, Gibbon,
  Zepf, Kar, Evans, and Bellei}}]{robinson2009relativistically}
\bibinfo{author}{\bibfnamefont{A.}~\bibnamefont{Robinson}},
  \bibinfo{author}{\bibfnamefont{P.}~\bibnamefont{Gibbon}},
  \bibinfo{author}{\bibfnamefont{M.}~\bibnamefont{Zepf}},
  \bibinfo{author}{\bibfnamefont{S.}~\bibnamefont{Kar}},
  \bibinfo{author}{\bibfnamefont{R.}~\bibnamefont{Evans}}, \bibnamefont{and}
  \bibinfo{author}{\bibfnamefont{C.}~\bibnamefont{Bellei}},
  \bibinfo{journal}{Plasma Physics and Controlled Fusion}
  \textbf{\bibinfo{volume}{51}}, \bibinfo{pages}{024004}
  (\bibinfo{year}{2009}).

\bibitem[{\citenamefont{Naumova et~al.}(2009)\citenamefont{Naumova, Schlegel,
  Tikhonchuk, Labaune, Sokolov, and Mourou}}]{naumova2009hole}
\bibinfo{author}{\bibfnamefont{N.}~\bibnamefont{Naumova}},
  \bibinfo{author}{\bibfnamefont{T.}~\bibnamefont{Schlegel}},
  \bibinfo{author}{\bibfnamefont{V.}~\bibnamefont{Tikhonchuk}},
  \bibinfo{author}{\bibfnamefont{C.}~\bibnamefont{Labaune}},
  \bibinfo{author}{\bibfnamefont{I.}~\bibnamefont{Sokolov}}, \bibnamefont{and}
  \bibinfo{author}{\bibfnamefont{G.}~\bibnamefont{Mourou}},
  \bibinfo{journal}{Physical Review Letters} \textbf{\bibinfo{volume}{102}},
  \bibinfo{pages}{025002} (\bibinfo{year}{2009}).

\bibitem[{\citenamefont{Weng et~al.}(2012)\citenamefont{Weng, Murakami, Mulser,
  and Sheng}}]{weng2012ultra}
\bibinfo{author}{\bibfnamefont{S.}~\bibnamefont{Weng}},
  \bibinfo{author}{\bibfnamefont{M.}~\bibnamefont{Murakami}},
  \bibinfo{author}{\bibfnamefont{P.}~\bibnamefont{Mulser}}, \bibnamefont{and}
  \bibinfo{author}{\bibfnamefont{Z.}~\bibnamefont{Sheng}},
  \bibinfo{journal}{New Journal of Physics} \textbf{\bibinfo{volume}{14}},
  \bibinfo{pages}{063026} (\bibinfo{year}{2012}).

\bibitem[{\citenamefont{Shen et~al.}(2007)\citenamefont{Shen, Li, Yu, and
  Cary}}]{shen2007bubble}
\bibinfo{author}{\bibfnamefont{B.}~\bibnamefont{Shen}},
  \bibinfo{author}{\bibfnamefont{Y.}~\bibnamefont{Li}},
  \bibinfo{author}{\bibfnamefont{M.}~\bibnamefont{Yu}}, \bibnamefont{and}
  \bibinfo{author}{\bibfnamefont{J.}~\bibnamefont{Cary}},
  \bibinfo{journal}{Physical Review E} \textbf{\bibinfo{volume}{76}},
  \bibinfo{pages}{055402} (\bibinfo{year}{2007}).

\bibitem[{\citenamefont{Gong et~al.}(2020)\citenamefont{Gong, Shou, Tang, Hu,
  Yu, Ma, Lin, and Yan}}]{gong2020proton}
\bibinfo{author}{\bibfnamefont{Z.}~\bibnamefont{Gong}},
  \bibinfo{author}{\bibfnamefont{Y.}~\bibnamefont{Shou}},
  \bibinfo{author}{\bibfnamefont{Y.}~\bibnamefont{Tang}},
  \bibinfo{author}{\bibfnamefont{R.}~\bibnamefont{Hu}},
  \bibinfo{author}{\bibfnamefont{J.}~\bibnamefont{Yu}},
  \bibinfo{author}{\bibfnamefont{W.}~\bibnamefont{Ma}},
  \bibinfo{author}{\bibfnamefont{C.}~\bibnamefont{Lin}}, \bibnamefont{and}
  \bibinfo{author}{\bibfnamefont{X.}~\bibnamefont{Yan}},
  \bibinfo{journal}{Physical Review E} \textbf{\bibinfo{volume}{102}},
  \bibinfo{pages}{013207} (\bibinfo{year}{2020}).

\bibitem[{\citenamefont{Thomas}(1927)}]{thomas1927kinematics}
\bibinfo{author}{\bibfnamefont{L.~H.} \bibnamefont{Thomas}},
  \bibinfo{journal}{The London, Edinburgh, and Dublin Philosophical Magazine
  and Journal of Science} \textbf{\bibinfo{volume}{3}}, \bibinfo{pages}{1}
  (\bibinfo{year}{1927}).

\bibitem[{\citenamefont{Bargmann et~al.}(1959)\citenamefont{Bargmann, Michel,
  and Telegdi}}]{bargmann1959precession}
\bibinfo{author}{\bibfnamefont{V.}~\bibnamefont{Bargmann}},
  \bibinfo{author}{\bibfnamefont{L.}~\bibnamefont{Michel}}, \bibnamefont{and}
  \bibinfo{author}{\bibfnamefont{V.}~\bibnamefont{Telegdi}},
  \bibinfo{journal}{Physical Review Letters} \textbf{\bibinfo{volume}{2}},
  \bibinfo{pages}{435} (\bibinfo{year}{1959}).

\bibitem[{\citenamefont{Pukhov and Meyer-ter Vehn}(1996)}]{pukhov1996_channel}
\bibinfo{author}{\bibfnamefont{A.}~\bibnamefont{Pukhov}} \bibnamefont{and}
  \bibinfo{author}{\bibfnamefont{J.}~\bibnamefont{Meyer-ter Vehn}},
  \bibinfo{journal}{Phys. Rev. Lett.} \textbf{\bibinfo{volume}{76}},
  \bibinfo{pages}{3975} (\bibinfo{year}{1996}).

\bibitem[{\citenamefont{Lasinski et~al.}(1999)\citenamefont{Lasinski, Langdon,
  Hatchett, Key, and Tabak}}]{lasinski1999particle}
\bibinfo{author}{\bibfnamefont{B.~F.} \bibnamefont{Lasinski}},
  \bibinfo{author}{\bibfnamefont{A.~B.} \bibnamefont{Langdon}},
  \bibinfo{author}{\bibfnamefont{S.~P.} \bibnamefont{Hatchett}},
  \bibinfo{author}{\bibfnamefont{M.~H.} \bibnamefont{Key}}, \bibnamefont{and}
  \bibinfo{author}{\bibfnamefont{M.}~\bibnamefont{Tabak}},
  \bibinfo{journal}{Physics of Plasmas} \textbf{\bibinfo{volume}{6}},
  \bibinfo{pages}{2041} (\bibinfo{year}{1999}).

\bibitem[{\citenamefont{Nakamura et~al.}(2010)\citenamefont{Nakamura, Bulanov,
  Esirkepov, and Kando}}]{nakamura2010_MVA_PRL}
\bibinfo{author}{\bibfnamefont{T.}~\bibnamefont{Nakamura}},
  \bibinfo{author}{\bibfnamefont{S.~V.} \bibnamefont{Bulanov}},
  \bibinfo{author}{\bibfnamefont{T.~Z.} \bibnamefont{Esirkepov}},
  \bibnamefont{and} \bibinfo{author}{\bibfnamefont{M.}~\bibnamefont{Kando}},
  \bibinfo{journal}{Physical review letters} \textbf{\bibinfo{volume}{105}},
  \bibinfo{pages}{135002} (\bibinfo{year}{2010}).

\bibitem[{\citenamefont{Liu et~al.}(2020)\citenamefont{Liu, Meyer-ter Vehn,
  Ruhl, and Zepf}}]{liu2020front}
\bibinfo{author}{\bibfnamefont{B.}~\bibnamefont{Liu}},
  \bibinfo{author}{\bibfnamefont{J.}~\bibnamefont{Meyer-ter Vehn}},
  \bibinfo{author}{\bibfnamefont{H.}~\bibnamefont{Ruhl}}, \bibnamefont{and}
  \bibinfo{author}{\bibfnamefont{M.}~\bibnamefont{Zepf}},
  \bibinfo{journal}{Plasma Physics and Controlled Fusion}
  \textbf{\bibinfo{volume}{62}}, \bibinfo{pages}{085014}
  (\bibinfo{year}{2020}).

\bibitem[{\citenamefont{Schlegel et~al.}(2009)\citenamefont{Schlegel, Naumova,
  Tikhonchuk, Labaune, Sokolov, and Mourou}}]{schlegel2009relativistic}
\bibinfo{author}{\bibfnamefont{T.}~\bibnamefont{Schlegel}},
  \bibinfo{author}{\bibfnamefont{N.}~\bibnamefont{Naumova}},
  \bibinfo{author}{\bibfnamefont{V.}~\bibnamefont{Tikhonchuk}},
  \bibinfo{author}{\bibfnamefont{C.}~\bibnamefont{Labaune}},
  \bibinfo{author}{\bibfnamefont{I.}~\bibnamefont{Sokolov}}, \bibnamefont{and}
  \bibinfo{author}{\bibfnamefont{G.}~\bibnamefont{Mourou}},
  \bibinfo{journal}{Physics of Plasmas} \textbf{\bibinfo{volume}{16}},
  \bibinfo{pages}{083103} (\bibinfo{year}{2009}).

\bibitem[{\citenamefont{Wan et~al.}(2020)\citenamefont{Wan, Andriyash, Lu,
  Mori, and Malka}}]{Wan_2020_PRL}
\bibinfo{author}{\bibfnamefont{Y.}~\bibnamefont{Wan}},
  \bibinfo{author}{\bibfnamefont{I.~A.} \bibnamefont{Andriyash}},
  \bibinfo{author}{\bibfnamefont{W.}~\bibnamefont{Lu}},
  \bibinfo{author}{\bibfnamefont{W.~B.} \bibnamefont{Mori}}, \bibnamefont{and}
  \bibinfo{author}{\bibfnamefont{V.}~\bibnamefont{Malka}},
  \bibinfo{journal}{Phys. Rev. Lett.} \textbf{\bibinfo{volume}{125}},
  \bibinfo{pages}{104801} (\bibinfo{year}{2020}).

\end{thebibliography}
\end{document}